\documentclass[useAMS,usenatbib]{mn2e}
\usepackage{amsfonts,amsmath,amssymb}
\usepackage{graphicx,epsfig}

\newcommand{\Msun}{\mbox{M$_\odot$}}

\journal{Preprint astro-ph/yymmnnn}

\title[UCD formation in Fornax]{Formation of ultra-compact dwarf
  galaxies: tests of the galaxy threshing scenario in Fornax}
\author[Thomas et al.]
{P.~A.~Thomas$^1$\thanks{Email: p.a.thomas@sussex.ac.uk},
 M.~J.~Drinkwater$^2$ and E.~Evstigneeva$^2$\\
$^1$Astronomy Centre, University of Sussex, Falmer, Brighton BN1 9QH\\
$^2$Department of Physics, University of Queensland, QLD 4072, Australia}

\date{\today}

\pagerange{\pageref{firstpage}--\pageref{lastpage}} \pubyear{2007}

\begin{document}

\maketitle

\label{firstpage}

\begin{abstract}

This paper investigates the possibility that UCD galaxies in the
Fornax cluster are formed by the threshing of nucleated, early-type
dwarf galaxies (hereafter dwarf galaxies).

Similar to the results of \citet{CPF06} for the Virgo cluster, we show
that the Fornax Cluster observations are consistent with a single
population in which all dwarfs are nucleated, with a ratio of nuclear
to total magnitude that varies slowly with magnitude.  Importantly,
the magnitude distribution of the UCD population is similar to that of
the dwarf nuclei in the Fornax cluster.

The joint population of UCDs and the dwarfs from which they may
originate is modelled and shown to be consistent with an NFW profile
with a characteristic radius of 5\,kpc.  Furthermore, a steady-state
dynamical model reproduces the known mass profile of Fornax.  However,
there are a number of peculiarities in the velocity dispersion data
that remain unexplained.

The simplest possible threshing model is tested, in which dwarf
galaxies move on orbits in a static cluster potential and are threshed
if they pass within a radius at which the tidal force from the cluster
exceeds the internal gravity at the core of their dark matter halo.  This
fails to reproduce the observed fraction of UCDs at radii greater than
30\,kpc from the core of Fornax. 

\end{abstract}

\begin{keywords}
galaxies: dwarf -- galaxies: formation -- methods: analytical 
\end{keywords}

\section{Introduction}

In recent years considerable evidence has accumulated that disruptive
processes play an important role in galaxy evolution as well as the
more dominant hierarchical merging. Observational evidence for these
disruptive processes is particularly evident in the dense environment
of galaxy clusters. The evidence includes populations of individual
intra-cluster objects such as planetary nebulae \citep{Arn04,FCJ04}
and red giant stars \citep{Dur02}, as well as the general diffuse
light now thought to make up a significant fraction of the total
stellar mass in clusters \citep*{FCJ04,FMM04,GZZ05,Zib05}.

In this paper we focus on a relatively new component of intra-cluster
space, ultra-compact dwarf (UCD) galaxies. These are compact systems
of old stars akin to globular clusters but they are 10--100 times more
luminous than Galactic globular clusters and they are located in
intra-cluster space between galaxies. The first UCDs were discovered
in the Fornax Cluster independently in studies of globular clusters
\citep*{Min98,HIV99} and in studies of compact dwarf galaxies
\citep{DJG00,PDG01}. The UCDs are unlike any known galaxies in terms
of luminosity, morphology and size \citep{DGH03}. Several hypotheses
have been suggested to explain the origin of UCDs ranging from them
being the high-luminosity end of a putative intra-cluster globular
cluster distribution, to being the evolved super star clusters formed
in galaxy merger events. In this paper we focus on the model that UCDs
are formed by the global tidal field of a cluster which can strip, or
``thresh'', the outer stellar envelopes of nucleated dwarf galaxies
(dE,Ns and dS0,Ns) as they pass repeatedly through the inner regions
of a cluster leaving just the bare nucleus to survive as a UCD
\citep*{BCD01,BCD03,GMK07}

The motivation for the current work is the subsequent discovery of a
larger population of fainter UCDs in the central region of the Fornax
Cluster \citep{DGC04,Gre08}. This sample of
60 UCDs is large enough to permit us to test several aspects of the
threshing hypothesis using a statistically significant sample. Our
focus will be to test simple aspects of the distributions of the UCD
and galaxy populations. An alternative approach based on the internal
properties of the UCDs is also in progress \citep[e.g.~][]{EGD07}.

Our basic premise for this paper is that if UCDs are descendants of
disrupted galaxies, then the UCD parent population can be modelled by
the combined {\em current} population of Fornax cluster UCDs and
dwarf galaxies.  We test whether the observed spatial and
velocity distributions of the two populations are consistent with this
hypothesis and conclude that they are.  We then model
the orbits of UCDs/galaxies drawn from this joint population to
determine what fraction of them pass close enough to the cluster
centre to lead to threshing.  The relative fraction UCDs to 
dwarfs seen at large radii in Fornax is inconsistent with this static
threshing model.

In Section~\ref{sec:data} we define the UCD and galaxy samples for our
analysis. In Section~\ref{sec:luminosity} we test if the luminosity
function of the UCDs is consistent with them having been drawn as
random sample from the nuclei of dwarf galaxies in the
cluster. Section~\ref{sec:dynamical} develops a dynamical model for
the joint population, and Section~\ref{sec:thresh} calculates the
fraction of threshed orbits at each radius.  Finally, in
Section~\ref{sec:conc}, we summarise our results and draw conclusions
about the plausibility  of the threshing hypothesis.

We adopt a distance of 20\,Mpc to the Fornax Cluster \citep*{DGC01}
corresponding to a distance modulus of 31.51 magnitudes.  In this
paper we are not concerned with late-type galaxies.  To avoid endless
repetition, we use the terms galaxy and dwarf to refer to early-type
objects only, as defined in Section~\ref{sec:early}.

\section{Data samples from the Fornax Cluster}
\label{sec:data}


\subsection{Early-type galaxy sample}
\label{sec:early}

The hypothesis that we test in this paper is that UCDs form from the
disruption of nucleated dwarf galaxies. Our authority for the
morphological classification of Fornax Cluster galaxies is the {\em
Fornax Cluster Catalog} \citep[FCC,][]{Fer89} which was based on
photographic data. The FCC lists some 291 galaxies as early types
(i.e.~not Sa-d, Sm or Im; we include spheroidal galaxies in our
sample). Of these, 103 are classified as nucleated. Recent {\em Hubble
Space Telescope} imaging results from the ACS Virgo Cluster Survey
\citep[hereafter CPF06]{CPF06} suggest that the frequency of
nucleation in early type galaxies is actually much higher than
suggested from the photographic ground-based surveys. Faint nuclei are
difficult to detect because they are washed out by atmospheric seeing
and the central regions of the brightest galaxies are saturated.
Notably, CPF06 suggest that potentially {\em all} dwarf galaxies may
contain nuclei.  We apply the CPF6 model to our Fornax data in
Section~\ref{sec:nucleated} and show that the observed fraction of
nucleated dwarfs as a function of magnitude is consistent with this
assumption.  Also, the spatial distributions of nucleated and
non-nucleated dwarfs, shown in Fig.~\ref{fig:compare}, are
indistinguishable.

For the purposes of this current work, therefore, we define the parent
galaxy sample to be all early-type dwarf galaxies listed as definite
or probable members in the FCC.  Where radial velocities are known, we
use these to define membership, otherwise we use the FCC membership
classifications. New radial velocities result in the removal of some
FCC-classified members and the inclusion of some FCC-classified
background galaxies now know to be members
\citep[e.g.~see][]{DGH01}. More recent radial velocity measurements
are taken from \citet*{KDG03}.  Where the classification is uncertain,
we have taken all galaxies fainter than $M_B=-14$ as dwarf; the
maximum magnitude for a normal galaxy is then $M_B=-16.3$ and the
minimum magnitude for a dwarf galaxy is $M_B=-17.8$.  A complete list
of the galaxies is given in Table~A1.

The galaxies in our sample have morphological classifications from the
FCC which can include a flag that they are nucleated. We use these
flags in our discussion below, but we emphasise that there is no HST
imaging for most of these galaxies, so we cannot tell with certainty
if a given galaxy is really nucleated. We instead adopt the general
result of CPF06 that all dwarf galaxies have nuclei, with a magnitude
that is related to that of the host galaxy (see
Section~\ref{sec:nucleated} below).

The dwarf galaxy sample that we use is effectively complete to a limit of
around $M_B=-13.5$ (see FCC). The velocity data used to confirm
cluster membership are complete for galaxies brighter than $M_B=-16$
and become 50 per cent complete at $M_B=-14.5$.

\subsection{UCD sample}

The UCDs were originally discovered as part of the all-object Fornax
Cluster Spectroscopic Survey \citep{DPJ00}. Although the original
survey measured all objects (to $b_J<19.8$), this was subsequently
extended in a selective search at fainter limits specifically designed
to find UCDs. The selected search used a slightly smaller field: a
radius limit of 0.9 degrees (314 kpc) around the central cluster
galaxy NGC~1399. The selected search also used a restricted colour
range of $b_J-r_F<1.7$ for objects with both $b_J$ and $r_F$ values
measured; for fainter objects with no $r_F$ values, no colour
criterion was applied. This colour selection served to remove Galactic
M-dwarf stars from the sample: no UCDs have colours in this range. The
completeness of the spectroscopic observations is given as a function
of magnitude in Table~\ref{tab:limits}.

The $b_J$ photographic APM magnitudes were converted to $m_B$
magnitudes by the approximate relation $m_B=b_J+0.20$ (based on the
\citealt{BlG82} relation of $b_J = B - 0.28(B-V)$ for an average
dwarf galaxy colour of $B-V=0.7$) so that $M_B = b_J +0.20 -31.51 =
b_J -31.31$.

\begin{table}
\caption{Spectroscopic completeness of the UCD sample.  Completeness,
$C$, is defined as the fraction of UCD targets for which redshifts
were measured. $N_{\rm UCD}$, in two bins of projected radius $r$, is the number of UCDs found in each magnitude
range. The actual number of UCDs can therefore by estimated as
$N_{\rm UCD}/C$.}
\label{tab:limits}
\begin{tabular}{ccccc}
\hline
$b_J$ range & $M_B$ range & $C$ & \multicolumn{2}{c}{$N_{\rm UCD}$} \\
            &             &     & $r<17.5$\,kpc & $r>17.5$\,kpc \\
\hline
16.0--20.5  & $-$15.3 to $-$10.8 &  0.94 & 0 & 21 \\
20.5--21.0  & $-$10.8 to $-$10.3 &  0.81 & 6 & 14 \\
21.0--21.5  & $-$10.3 to $-$9.8  &  0.35 & 5 & 14 \\
\hline
\end{tabular}
\end{table}

The spatial locations of the UCDs are far from circularly symmetric
about the centre of Fornax but tend to lie in a band running from
North East to South West \citep[see fig.~2 in][]{Gre08}.  This
presumably reflects the infall pattern onto the cluster.  Provided
that the distribution is relaxed then this will not affect the
dynamical modelling; however it may confuse the relation between the
true three-dimensional positions and velocities and the observed
ones. For the purposes of modelling in this paper, we assume a
spherically-symmetric distribution.

\subsection{Joint sample selection}
\label{sec:joint}

According to our central hypothesis, there was an original parent
population of dwarf galaxies, some of which were subsequently
disrupted to form UCDs.  Unfortunately the selection effects are
different for the two sub-populations and so we need to use different
samples for different parts of our analysis.  This will be described
at the beginning of each relevant section.  Here we make a few
general comments on the relative spatial extent of the dwarf and UCD
samples.

The FCC is a wide-field survey.  It covers a rectangular region with a
largest inscribed circle that extends to a radius of 3 degrees (1.05
Mpc) from the cluster centre. Our main UCD sample is limited to a
smaller region defined by a maximum radius of 0.9 degrees (314 kpc)
from the cluster centre.  We have modelled the density
distribution and estimate that there may be up to 6 missing UCDs at
larger radii (although, for the brighter UCDs, two additional regions
extending to a radius of 3 degrees have been surveyed and no UCDs were
found).  Adding 6 extra UCDs with the appropriate density distribution
makes very little difference to the modelling of the spatial
distribution of the joint UCD plus dwarf population in
Section~\ref{sec:spatial} below.

\begin{figure}
\psfig{file=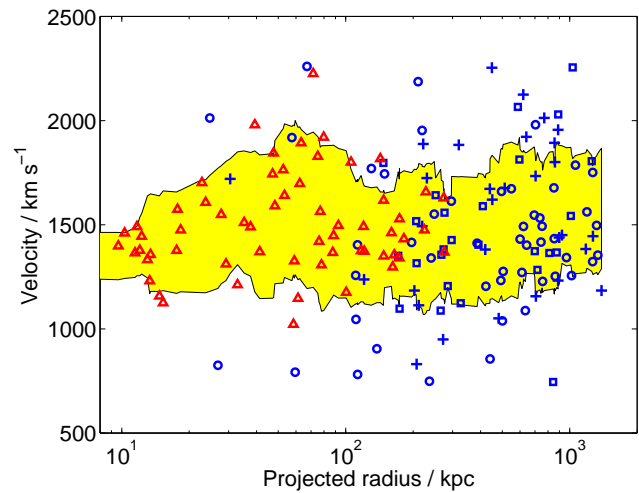,width=\linewidth}
\caption{A comparison of the UCD and galaxy populations. The radial
  velocities are plotted as a function of projected radius. UCDs are
  shown as red triangles, normal galaxies as blue squares, and dwarfs
  brighter and fainter than $M_B=-15.0$ as blue circles and crosses,
  respectively.  The yellow shaded area shows a running mean of the
  1-sigma velocity dispersion.}
\label{fig:velocities}
\end{figure}

For the UCDs there is also a need to exclude those at very small radii
from the central cluster galaxy, NGC~1399. The distribution of UCD
radial velocities shown in Fig.~\ref{fig:velocities} shows a trend to
smaller velocities (and velocity dispersion) at low radius.  The inner
UCDs are clearly moving in the galactic and not the cluster potential
and could be considered as bright globular clusters attached to
NGC1399.  The choice of where to draw the dividing line between
galactic and intracluster UCDs is somewhat arbitrary.  We cut at 3
arcmin (17.5\,kpc) which excludes 11 UCDs from our sample (see
Table~\ref{tab:limits}), including the two relatively low-velocity
UCDs seen in the figure at a radius and velocity of approximately
15\,kpc and 1140\,km\,s$^{-1}$, respectively.  (Including these two in
our analysis makes little difference to the results and would leave
the velocity dispersion of the excluded clusters as formally zero once
the velocity errors have been accounted for.)

For reasons that we shall describe later in Section~\ref{sec:UCDs}, we
divide the dwarf population into two.  ``Bright dwarfs'' with
$M_B<-15.0$ are those that correspond to the progenitors of UCDs in
our model, whereas ``faint dwarfs'' would give UCDs that fall below
our magnitude limit.  Table~\ref{tab:numbers} gives the number of
galaxies of each type in different radial bins.

\begin{table}
\caption{Number of Fornax galaxies of different types in different
  annular bins centred on NGC\,1399.  The division between bright and
  faint dwarfs is taken as $M_B=-15.0$.}
\label{tab:numbers}
\begin{tabular}{lccc}
\hline
Sample & \multicolumn{3}{c}{Radial range in kpc} \\ 
       & 17.5--314& 314--1\,050& 1\,050--1\,500\\
\hline
UCDs&            49&   0&  0\\
Normal galaxies& 13&  13&  4\\
Bright dwarfs&   11&  22&  5\\
Faint dwarfs&    65& 129& 28\\
\hline
\end{tabular}
\end{table}

\section{Comparison of luminosity distributions}
\label{sec:luminosity}

In this section we develop a unified model for nucleated and
non-nucleated dwarf galaxies whereby all galaxies have nuclei but only
a fraction of these are bright enough to be detected and classified as
such in the FCC.  We then go on to compare the predicted luminosity
function of nuclei with that of UCDs.  As we are interested only in
the shape of the magnitude distributions, we use the full samples of
dwarfs and UCDs even though two extend over different spatial regions.

\subsection{Early-type nuclei}
\label{sec:nucleated}

An important property of the parent galaxies is the luminosity
distribution of the galaxy nuclei as these will be compared to the UCD
luminosities.  We cannot directly measure the luminosities of galaxy
nuclei in the Fornax Cluster because most do not have high-resolution
HST imaging.  Instead we take a statistical approach: we assume that
{\em all} dwarf galaxies host nuclei and infer the nuclear
luminosities from the total galaxy luminosities.

CPF06 measured nuclear luminosities for 51 dwarf galaxies in the Virgo
Cluster. They confirmed previous suggestions that the nuclear
luminosities increase with the galaxy luminosity.  They modelled this
relation as both a fixed offset between the nuclear and total
magnitudes, $g'_{\rm nuc}= g'_{\rm gal} + (6.25 \pm 0.21)$, and an
offset slowly varying with magnitude, $g'_{\rm nuc}= (0.90 \pm 0.18)
g'_{\rm gal} + (7.59 \pm 2.50)$.  We note that our galaxy sample
extends to much fainter magnitudes than did the Virgo sample studied
by CPF06, and so we will have to extrapolate their relation.  We
therefore allow the slope of the relation to vary, but require that it
go through the mid point of the CPF06 data ($g'_{\rm gal},g'_{\rm
nuc}$ = $13.40$, 19.65) defined by the crossing point of their two
relations.

We model the scatter in the relationship by adding a random normal
variable with a mean of zero and a standard deviation of 1.5 to the
derived nuclear magnitude. The standard deviation was inferred from
the scatter about the fixed-slope fit of CPF06 (their equation 15).

To convert the ACS $g'$ photometry to absolute magnitudes we first use
the mean value of $B_T-g=0.30$ for the ACS dwarf galaxies to convert
$g$ magnitudes to $B_T$. We then apply the distance modulus of 31.09
magnitudes quoted by CPF06, obtaining $M_B=g'-30.79$.

To constrain the slope of the $g'_{\rm nuc}-g'_{\rm gal}$ relation, we
require that it predicts the correct distribution of galaxies that we
would expect to have been classified as nucleated in the photographic
FCC survey.  For each galaxy, we predict its nuclear luminosity as
above, then we classify it as nucleated if the nucleus is brighter
than the point-source detection limit on the photographic plate
(approximately $B_T = 22.6$ or $M_B=-8.9$ for the FCC; H.\ Ferguson, private communication).

\begin{figure}
\psfig{file=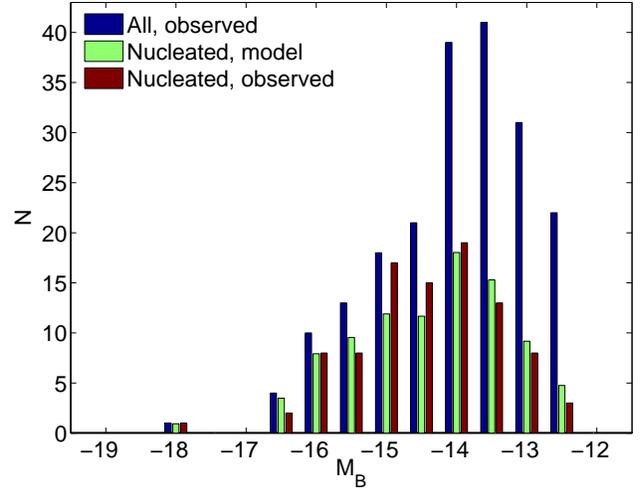,width=\linewidth}
\caption{The distribution of dwarf galaxies classified as
  ``nucleated''.  The leftmost, blue bars show the observed magnitudes of
  all Fornax dwarfs.  The middle, green bars show only those whose
  predicted nuclear magnitudes would be greater than $-8.9$ according
  to the model developed in the text.  Finally, the rightmost, brown bars
  show the actual magnitude distribution of nucleated dwarfs in Fornax.}
\label{fig:compare_den}
\end{figure}

The results, shown in Fig.~\ref{fig:compare_den}, show that this model
can nicely predict the luminosity distribution of the early-type
galaxies that are classified as nucleated in the photographic FCC
survey.  To produce the figure, we used a slope of 0.7, smaller than
the best fit of CPF06.\footnote{It is not clear as to whether the
CPF06 data will accept a slope of 0.7: we can also get an acceptable
fit if we use a slope of 0.9 for the relation, provided that the
magnitude limit for point-source detection is raised to $-8.5$.\\ If
we extend our analysis to include normal galaxies then we predict far
more nucleated galaxies than are observed.  This seems to be in
disagreement with the results from Virgo; however it is possible that
brightest nuclei would be saturated on the photographic plate and so
hard to detect: we note that CPF06 find many more nuclei in bright
galaxies with $M_B\leq-17.4$, than did photographic surveys.}  The
nuclear-to-total magnitude relation becomes
\begin{equation}
M_{B{\rm nuc}}+11.14= 0.7 (M_{B{\rm gal}}+17.39).
\label{eq:nucmag}
\end{equation}
The leftmost, blue bars in the figure show the magnitude distribution
of all dwarfs.  We averaged over 100 realisations of the scatter in
the relation to obtain the prediction for observable, nucleated dwarfs
shown in the green, middle bars of the figure.  These are
statistically indistinguishable from the actual number of dwarfs
classified as nucleated, shown in the rightmost, brown bars.

We show in Section~\ref{sec:spatial} below that the spatial
distributions of the nucleated and non-nucleated dwarfs are identical,
thus lending further support to the hypothesis that the presence of a
detectable nucleus is the only difference between them.

\subsection{UCDs}
\label{sec:UCDs}

\begin{figure}
\psfig{file=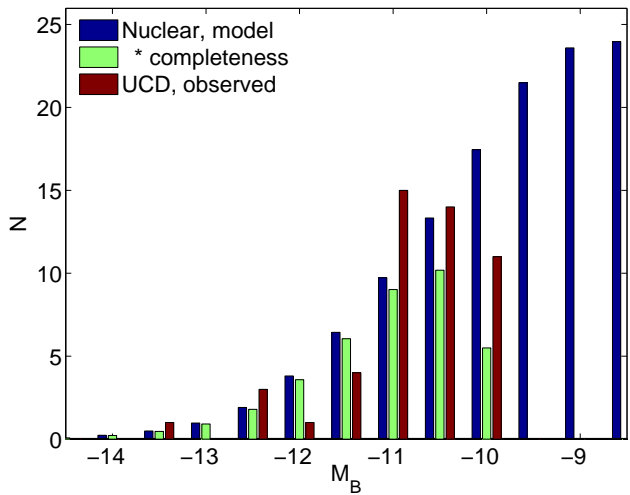,width=\linewidth}
\caption{The leftmost, blue bars show the predicted distribution of
  nuclear magnitudes for Fornax dwarfs, averaged over 100
  realisations: the middle, green bars multiply this by the
  completeness factor for UCD observations.  If the threshing
  hypothesis is correct then this should be proportional to the rightmost,
  brown bars that show the observed distribution of UCD magnitudes.}
\label{fig:compare_ucd}
\end{figure}

From the observed dwarf population, we can now predict the
distribution of nuclear magnitudes. If we assume that the threshing
process is independent of galactic (and nuclear) luminosity, then
these should have the same {\em shape} of distribution as the
UCDs. Furthermore, the relative normalisation should tell us what
fraction of the dwarfs have been threshed. The predicted and actual
UCD distributions are given in Fig~\ref{fig:compare_ucd}. Note that
the predicted numbers from our model have been scaled down by the
completeness values in Table~\ref{tab:limits} to allow for the
fraction of unmeasured objects.

The figure shows that the predicted luminosity distribution of UCDs is
not perfect.  The model seems to give an excess of UCDs brighter
than $M_B=-11.25$ as compared to fainter ones.  It is hard to assess
the significance of this: given the relatively small number of objects
and the uncertainties in the relationship between galactic and nuclear
magnitudes, it is probably acceptable.

The model predicts that 38 dwarfs should have nuclei that correspond
to observable UCDs.  This motivates our selection of $M_B=-15.0$ as
the dividing line between bright and faint dwarfs, as this gives 38
bright dwarfs.  Without scatter, equation~\ref{eq:nucmag} would have
predicted a brighter limit, $M_B\approx-15.8$, but the greater number
of faint galaxies biases things towards fainter magnitudes.  Of these
38 dwarfs, only 11 lie within 314\,kpc.  Thus the model predicts that
the vast majority of dwarfs within this region are likely to be
threshed.  Even when averaged over the whole sample, more than half
the dwarf population should be threshed.

\section{A dynamical model for the joint dwarf/UCD population}
\label{sec:dynamical}

This section constructs a model of the three-dimensional density
distribution of the joint dwarf/UCD population in the cluster. There
will turn out to be some degeneracy in the models which we will
attempt to constrain by matching them to observed mass models for
Fornax.  

When comparing dwarfs and UCDs, we restrict our attention to the
bright dwarfs, $M_B<-15.0$, as described in the previous section.  We
will show that the joint UCD + bright dwarf population is well-fit by
an NFW model \citep*{NFW97} in dynamical equilibrium in the cluster
potential.

\subsection{Spatial distributions}
\label{sec:spatial}

\begin{figure}
\psfig{file=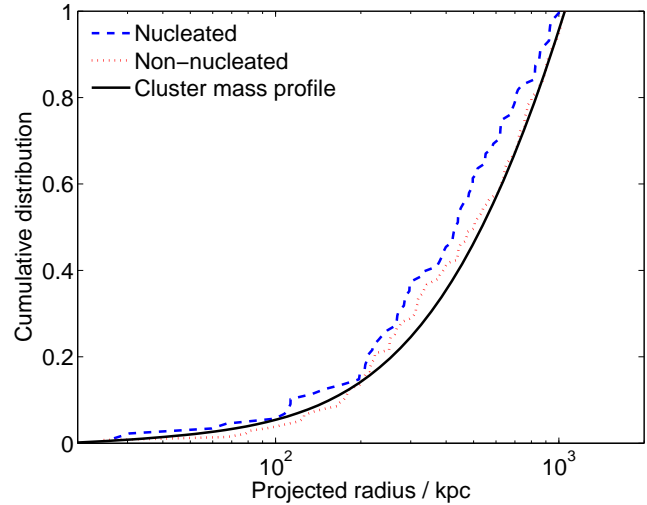,width=8.6cm}
\caption{The cumulative numbers of nucleated (dashed, blue) and
  non-nucleated (dotted, red) dwarf galaxies in Fornax as a
  function of projected radius between 17.5 and 1050\,kpc.  Shown also
  as a solid, black line is the cluster mass profile described in
  Section~\ref{sec:massprof}.}
\label{fig:compare}
\end{figure}

We first compare the spatial distribution of nucleated and
non-nucleated dwarfs.  If we plot all the dwarf galaxies together, as
in Fig.~\ref{fig:compare}, the nucleated dwarfs are shown to be very
slightly more centrally concentrated. This is in the same sense as was
originally reported for the dwarf galaxies of the Virgo \citep*{BTS87}
and Fornax \citep{FeS89} Clusters, but for our present sample the
difference is not significant (the Kolmogorov--Smirnov test returns a
probability that the two are drawn from the same distribution of
0.2). As expected from our discussion above, the nucleated dwarfs are
significantly more luminous on average than the non-nucleated dwarfs:
the difference in radial distributions is actually a luminosity
bias. If, instead, we compare the distributions of dwarfs of the same
luminosity ($M_B>-15$), then the difference is removed entirely.  This
observation strengthens the hypothesis of the previous section, that
dwarfs classified as nucleated or non-nucleated may differ only in the
detectability of their central nucleus.

Also shown in the figure is the observed mass profile of the Fornax
cluster, as described in Section~\ref{sec:massprof}. The cumulative
number density profile of the dwarfs matches that of the cluster mass
profile very well and shows no evidence of dwarf galaxy disruption
near the cluster core.

\begin{figure}
\psfig{file=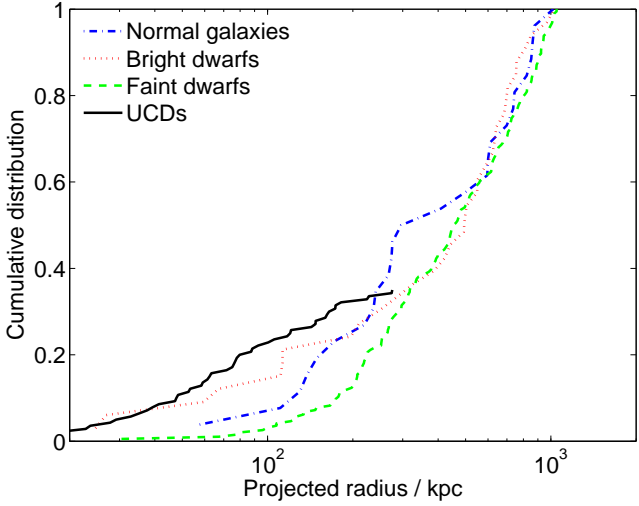,width=8.6cm}
\caption{The cumulative numbers of normal galaxies (dash-dotted,
  blue), bright dwarfs (dotted, red), faint dwarfs (dashed, green) and
  UCDs (solid, black) in Fornax as a function of projected radius
  between 17.5 and 1050\,kpc.  We have adjusted the normalisation of
  the UCD curve to match that of the dwarfs at 314\,kpc.} 
\label{fig:compare_denucd}
\end{figure}

Next, in Fig.~\ref{fig:compare_denucd} we compare the radial
distributions of normal galaxies, bright and faint dwarfs and UCDs.
Because the UCD survey extends only out to 314\,kpc, we have adjusted
the normalisation of the cumulative distribution to match that of the
dwarfs at this radius.

It is immediately obvious that the different populations show
different degrees of central concentration.  Notably, within 314\,kpc,
the radial distribution of the faint dwarfs is significantly more
extended than that of both the bright dwarfs and the UCDs.  At first
glance, this appears to be at odds with the threshing model developed
below (Section~\ref{sec:threshradii}). The model suggests that faint
dwarfs are more compact and therefore less likely to be threshed than
bright ones, but we have not looked for UCDs at magnitudes
corresponding to the faint dwarfs so we cannot test the number that
have been threshed. Conversely, we do not see a significant difference
between the distributions of UCDs and bright dwarfs, although we would
expect the UCDs to be more centrally-concentrated than the (surviving)
dwarf galaxies according to our model. In this case, the relatively
small number of objects involved may explain why the difference is not
significant.

Outside 314\,kpc the distributions of bright and faint dwarfs are
indistinguishable.  There are hints that the UCD distribution is
flattening between 200 and 314\,kpc and no UCDs have been detected in
(incomplete) observations in a few fields beyond this radius.  For the
purposes of the modelling that follows, we therefore assume that there
are no UCDs with cluster-centric radii exceeding 314\,kpc.  If there
are any, the number density of dwarfs rises so rapidly in this region
that the latter would dominate anyway.

We plot the radial distribution of normal galaxies just for interest.
The numbers are so few that it is formally indistinguishable from
either the bright or faint dwarf population.  We note, however, that
it is significantly less centrally-concentrated within 314\,kpc than
the UCD population.

Assuming a spherically-symmetric distribution dependent only on radius,
$r$, we model the density, $\rho(r)$, with profiles of the form
\begin{equation}
\rho=\frac{\rho_0}{x\,(1+x)^{s-1}},
\label{eq:density}
\end{equation}
where $x=r/a$, and $a$ and $s$ are fitting parameters.  (We fit only
for the shape of the distribution: the normalisation $\rho_0$ can be
chosen so as to match the correct number of objects.)  We project each
distribution onto the sky and then compare the predicted cumulative
mass profile as a function of radius to the observed distribution,
using a Kolmogorov--Smirnov test.

\begin{figure}
\psfig{file=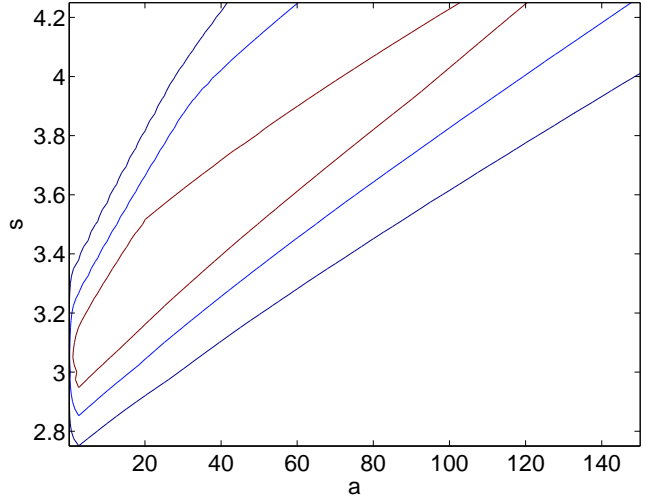,width=8.6cm}
\caption{The 1-, 2-, and 3-sigma range of allowable parameters in the
  fit to the joint UCD plus bright dwarf population.}
\label{fig:probmock}
\end{figure}

\begin{figure}
\psfig{file=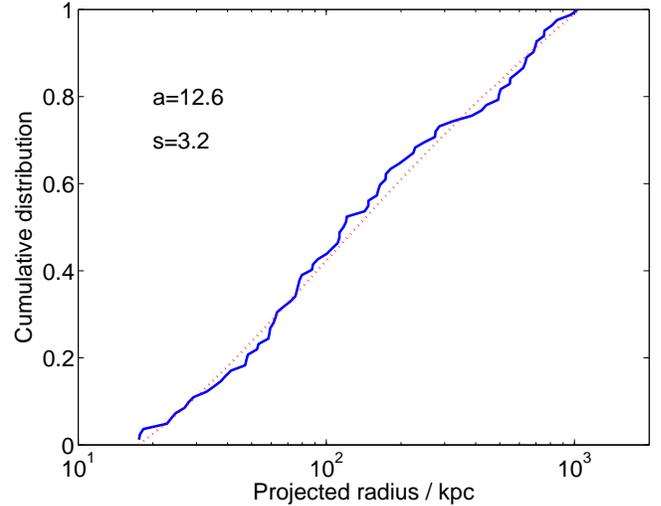,width=8.6cm}
\caption{The best-fitting cumulative profile of the total number of
  UCD+bright dwarf galaxies as a function of projected radius: (solid, blue)
  data, (dotted, red) model.}
\label{fig:bestmock}
\end{figure}

Fig.~\ref{fig:probmock} shows the allowable range of parameters and
Fig.~\ref{fig:bestmock} shows the best-fit model, although there is a
strong degeneracy between $a$ and $s$ such that a wide variety of fits
are acceptable.  We will show results for $s=3.0$, $a=5$\,kpc and for
$s=4.0$, $a=90$\,kpc; both lead to very similar conclusions.

\subsection{Velocities}

The velocity dispersions for different sub-samples of Fornax UCDs and
galaxies are shown in Table~\ref{tab:sigmas}.  To determine the population
velocity dispersions, $\sigma$, we used the following formula:
\begin{equation}
\sigma^2\sum_i w_i=\sum_i w_i \big( (v_i-\bar{v})^2 -
\sigma_{e,i}^2\big),
\end{equation}
where $v_i$ and $\sigma_{e,i}$ are the observed velocities and their
rms measurement errors, respectively, $\bar{v}$ is the mean velocity
for the full sample of all galaxies plus UCDs outside 17.5\,kpc,
\begin{equation}
\bar{v}\sum_i w_i=\sum_i w_i v_i,
\end{equation}
and
\begin{equation}
w_i={1\over \sigma^2+\sigma_{e,i}^2}
\end{equation}
are weights chosen so as to maximise the information in the
data.\footnote{We do not have any formal proof of this but note that
the weights are equal when $\sigma_{e,i}\ll\sigma$ and tend to the
known optimal weighting $w_i\propto 1/\sigma_{e,i}^2$ when
$\sigma_{e,i}\gg\sigma$ \citep{Irw42}.  A similar, but not identical
expression is given by \citet{PrM93}.}

\begin{table}
\caption{Velocity dispersions for different subsamples of the UCD and
  galaxy populations in Fornax.  The completeness of the velocity data
  and be found by comparing the numbers in this table with those in
  Table~\ref{tab:numbers}, but basically it is high except for the
  faint dwarfs.  When calculating velocity dispersions for the
  different sub-samples, we have used the mean for the full sample of
  all galaxies (normal, dwarf and UCD) with radii greater than
  17.5\,kpc, $\bar{v}=1491$\,km\,s$^{-1}$.  The final column shows the
  rms error in the velocity dispersion measurements determined by
  bootstrap resampling 1000 times.}
\label{tab:sigmas}
\begin{tabular}{lcccc}
\hline
Sample & Number & \multicolumn{2}{c}{$\sigma/$km\,s$^{-1}$}& error\\
       &        & raw & corrected\\
\hline
Full                         & 154& 321& 316& 19\\
UCDs + bright dwarfs         &  82& 316& 310& 27\\
\\
Normal                       &  29& 281& 280& 40\\
Bright dwarfs                &  33& 401& 399& 42\\
Faint dwarfs                 &  43& 356& 350& 31\\
UCDs                         &  49& 235& 224& 29\\
\\
UCDs, 17.5\,$<R/$kpc\,$<$\,76 & 24& 285& 276& 42\\
UCDs, \ 76\,$<R/$kpc\,$<$\,314& 25& 174& 159& 28\\
\hline
\end{tabular}
\end{table}

The low velocity dispersion of UCDs as compared to other galaxies is
expected in the threshing model, because the UCDs are more
centrally-concentrated in the cluster potential (i.e.\ have a steeper
density profile) -- unfortunately there are too few UCDs to quantify
this.  However, the table shows a number of other features that are
hard to explain.

Firstly, why is the velocity dispersion of normal galaxies so much
smaller than that of dwarfs, and especially bright dwarfs, given that
the two have similar radial distributions within the cluster?  In
\citet{DGC01} this difference was interpreted as indicating that the
dwarf galaxies were an unrelaxed, infalling, population. In this paper
we are assuming that all galaxies (including UCDs) are relaxed: an
alternative explanation is that many of the dwarfs may be orbiting in
bound subhalos, with normal galaxies located at their centres.

Secondly, the line-of-sight velocity dispersion for UCDs is
significantly higher at small radii than at large ones.  Some
difference of this kind would be expected if the UCDs are on
preferentially radial orbits.  Defining the velocity anisotropy
parameter as $\beta=1-\sigma_{\rm t}^2/\sigma_{\rm r}^2$, where
$\sigma_r$ is the radial velocity dispersion with resepct to the
cluster centre and $\sigma_t$ the tangential one, then this would
correspond to $\beta>0$.  Unfortunately, the expected variation,
calculated in Appendix~\ref{sec:losvel}, is much too small to explain
the observations.  The observed decline in velocity dispersion between
the inner and outer bin is 1:0.58.  Even if we allow each measurement
to move up to 1-sigma towards agreement (with probability less than 3
per cent), then the ratio remains 1:0.80.  This can only be explained
with $\beta=1$, corresponding to purely radial orbits.  The
explanation for this discrepancy may be related to the non-uniform
distribution of UCDs within the Fornax cluster.  If the outer UCDs
have orbits that are preferentially moving perpendicular to the
line-of-sight, then that would explain the effect.

Despite these uncertainties, we will model the joint UCD plus bright
dwarf population as if it is relaxed.  As we show in the next section,
this provides a marginally acceptable fit to the known mass
distribution in the Fornax cluster.

\subsection{Cluster mass profile}
\label{sec:massprof}

The mass of Fornax has been investigated using a number of different
techniques that probe different radial locations. \citet{RDG04}
look at the dynamics of the globular cluster population around
NGC\,1399.  They find that
\begin{equation}
{M\over\Msun}\approx 4.5\times10^{10}{r\over{\rm kpc}}
\end{equation}
for $r\lesssim20\,$kpc and increases in a similar vein to about twice
this radius.  This agrees with the ASCA observations of \citet{IEF96}
and the ROSAT observations of \citet{JSF97}.  The two X-ray papers
give different mass profiles at larger radii, but agree on a mass of
$10^{13}\Msun$ within 200\,kpc.  Finally, \citet{DGC01} have used the
shape of the velocity cusp to determine a mass for the cluster as a
whole of approximately $6\times10^{13}\Msun$ within 1\,Mpc.

\begin{figure}
\psfig{file=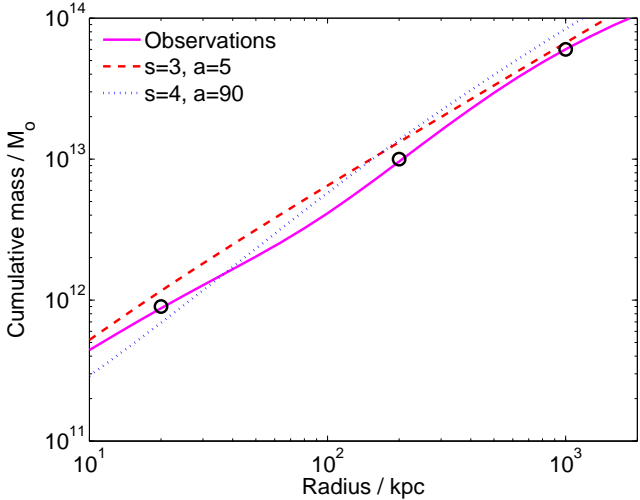,width=8.6cm}
\caption{A comparison of the observed and modelled cluster mass
  profiles for the case of $\beta=0$: observations (black circles and
  solid, magenta line), $s=3$, $a=5$\,kpc (dashed, red line); $s=4$,
  $a=90$\,kpc (dotted, blue line).}
\label{fig:massprof}
\end{figure}

We combine all these estimates into a density/mass model consisting of an
inner truncated isothermal sphere centred on NGC\,1399, plus a cluster
NFW potential:
\begin{equation}
\rho={\rho_{\rm BGC,0}\over x_{\rm BCG}^2\,(1+x_{\rm BCG}^2)}
     +{\rho_{\rm clus,0}\over x_{\rm clus}\,(1+x_{\rm clus})^2};
\label{eq:denprof}
\end{equation}
\begin{eqnarray}
M_r&=&M_{\rm BCG}\,{2\over\pi}\,\arctan x_{\rm BCG}\nonumber\\
&+&M_{\rm clus,0}\,\left[\ln(1+x_{\rm clus})-{x_{\rm clus}\over
  1+x_{\rm clus}}\right],
\label{eq:massprof}
\end{eqnarray}
where 
$M_{\rm BCG}=2\pi^2a_{\rm BCG}^3\,\rho_{\rm BCG,0}=2.0\times10^{12}\Msun$; 
$a_{\rm BCG}=30$\,kpc;
$M_{\rm clus,0}=4\pi\rho_{\rm clus,0}a_{\rm clus}^3=1.1\times10^{14}\Msun$;
$a_{\rm clus}=400$\,kpc.
Here $\rho(r)$ is the density at clustocentric radius $r$, $M_r$ is
the mass contained within radius $r$, $x_{\rm BCG}=r/a_{\rm BCG}$ and
$x_{\rm clus}=r/a_{\rm clus}$.
Given the uncertainties in the observations, any other model that has
$M\propto r$ in the centre, and that passes through the other mass
points mentioned above, would be equally acceptable.  The
observational constraints are shown in Fig.~\ref{fig:massprof} as
black circles, and the model as a solid, magenta line.

If the joint UCD plus bright dwarf population is to be at rest in the
cluster then it must satisfy the Jeans equation:
\begin{equation}
{1\over\rho}\,{{\rm d}\rho\sigma_r^2\over{\rm d}r}+{2\beta\sigma_r^2\over r}
=-{GM_r\over r^2}.
\label{eq:jeans}
\end{equation}
We can use this in two ways: to predict the mass distribution, given
our dynamical model for the population, or to predict the velocity
dispersion profile for the given observed mass profile.

Fig.~\ref{fig:massprof} shows a comparison between the observed
mass profile and that predicted by two of the acceptable density
models with constant velocity dispersion of 310\,km\,s$^{-1}$ and
isotropic velocity dispersion tensors, $\beta=0$.   
We have also tried models with $\beta>0$.  This makes very little
difference to the $s=3$ prediction but substantially worsens the $s=4$
fit to the data at small radii.

\begin{figure}
\psfig{file=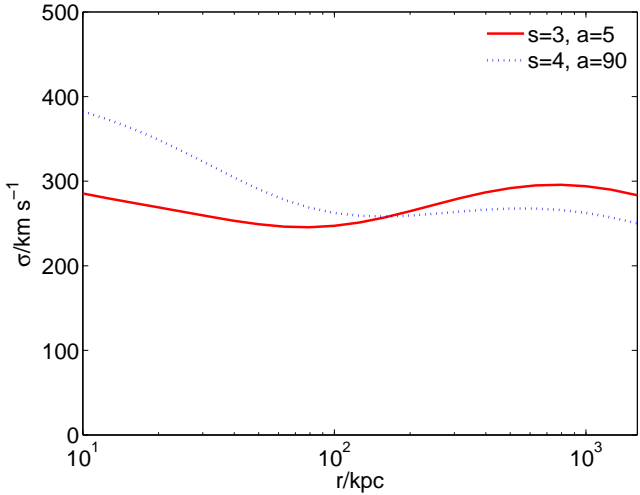,width=8.6cm}
\caption{The predicted isotropic velocity dispersion profiles for the
  mass model given in equation~\ref{eq:massprof} and for density
  profiles with parameters $s=3$, $a=5$\,kpc (solid, red) and $s=4$,
  $a=90$\,kpc (dotted, blue).}
\label{fig:sigmaprof}
\end{figure}

Reversing this procedure, Fig.~\ref{fig:sigmaprof} gives the
predicted velocity dispersion profile for a given density profile and
observed cluster mass distribution.  In making this prediction, we
have taken the approximation that the logarithmic gradient in the
velocity dispersion is small compared to that of the density.
Once again, taking $\beta$ to be greater than zero makes little
difference to the $s=3$ prediction, but worsens the $s=4$ one, giving
higher predicted velocity dispersion at small radii.

In both these plots, the $s=3$ curve provides the closer fit to the
data.  That it does not match every wiggle in the mass profile in
Fig.~\ref{fig:massprof} is not surprising given that the latter is
somewhat arbitrary and that we have not allowed the velocity
dispersion to vary with radius.  The normalisation is a little too
high: lowering the velocity dispersion to 283\,km\,s$^{-1}$ would
provide a very good fit to the mass profile.  Given that this is only
1-sigma away from the measured value in Table~\ref{tab:sigmas}, we
regard this as marginally acceptable.

In Fig.~\ref{fig:sigmaprof} it may seem at first sight that the
decline in velocity dispersion away from the core of the cluster
mimics that seen in the UCD observations.  However, a closer inspection
reveals that the minimum in velocity dispersion seen in this plot lies
at too small a radius and that by the edge of the UCD observations at
around 300\,kpc the velocity dispersion has risen to its central
value.

We conclude that a joint density profile for the combined UCD plus
bright dwarf population of the form of equation~\ref{eq:density} with
$s=3$ and $a=5$\,kpc provides an acceptable fit both to the
observed UCD plus dwarf population, in agreement with previous mass
estimates of the Fornax cluster.

\section{A static model of galaxy threshing}
\label{sec:thresh}

In this section, we investigate the simplest threshing scenario in
which dwarfs galaxies orbit within the present-day Fornax cluster and
are threshed if they pass close to the cluster core.  We show that
there are too many UCD galaxies at large radii for this model to be
viable.  We conclude that in any threshing model, disruption must
occur near the cores of smaller subclumps, prior to cluster formation.

\subsection{Threshing radii}
\label{sec:threshradii}

To estimate the fraction of dwarf orbits at a given radius which lead
to threshing, we calculate the probability for a galaxy with initial
projected clustocentric radius and line-of-sight velocity to have
$R_{\rm min} < R_{\rm th}$.  Here $R_{\rm min}$ is the minimum
distance from the cluster center during its orbit and $R_{\rm th}$ is
the radius within which the stellar envelope of a nucleated dwarf can
be removed by the tidal field of a cluster.

To determine $R_{\rm th}$ for a dwarf orbiting a cluster we assume
that $R_{\rm th}$ is the distance from the cluster center at which the
tidal force of the cluster equals the self-gravitational force of the
dark-matter halo in the inner regions of the dwarf galaxy.  This
occurs at the clustocentric radius for which
\begin{equation}
{GM_{\rm dm}\over r_{\rm dm}^2}=r_{\rm dm}\,
\left|{d\over dr}\left(GM_{\rm clus}\over r^2\right)\right|,
\label{eq:tidal}
\end{equation}
where $M_{\rm dm}$ is the dwarf halo mass within radius $r_{\rm dm}$
and $M_{\rm clus}$ is the cluster mass profile from
equation~\ref{eq:massprof}.

For the dark matter distribution in dwarf galaxies we use a profile
proposed by \citet{Bur95}:
\begin{equation}
\rho_{\rm dm}={\rho_{{\rm dm},0}\over(1+x_{\rm dm})(1+x_{\rm dm}^2)},
\end{equation}
where $x_{\rm dm}=r_{\rm dm}/a_{\rm dm}$, and $\rho_{{\rm dm},0}$ and $a_{\rm
dm}$ are the central dark matter density and the core (scale) radius,
respectively.  
This has an extended, constant-density central region within which
\begin{equation}
{M_{\rm dm}\over r_{\rm dm}^3}\approx 4.19\rho_{{\rm dm},0},
\end{equation}
so that $M_{\rm dm}$ and $r_{\rm dm}$ cancel in
equation~\ref{eq:tidal} leaving only a dependence upon $\rho_{{\rm dm},0}$.
We note that the central mass profiles of dwarf galaxies are not very
well known and could be more concentrated than assumed here.  Were
that to be the case, then the threshing radii would be reduced.

\citet{Bur95} gives observed scaling relations between $\rho_{{\rm
dm},0}$, $a_{\rm dm}$ and the circular velocity, $v_{{\rm dm},0}$, at
the core radius.  At that radius the velocity dispersion (assumed
isotropic) is approximately $\sigma\approx v_{{\rm dm},0}/\sqrt{2}$
and we assume that this is close to the observed value for the dwarf
as a whole.  The relevant relation is then:
\begin{equation}
{\rho_{{\rm dm},0}\over\Msun\,{\rm pc}^{-3}}\approx0.56
\left(\sigma\over{\rm km\,s}^{-1}\right)^{-1}.
\label{eq:rhodmobs}
\end{equation}

For each UCD, we determine the most likely magnitude of its precursor
dwarf using the relation of equation~\ref{eq:nucmag}.  We then use the
observed relationship from \citet{GGM03} to relate the magnitude to
velocity dispersion,
\begin{equation}
\log\left(\sigma\over{\rm km\,s}^{-1}\right) = 0.42 - 0.07\,M_B.
\end{equation}

\begin{figure}
\psfig{file=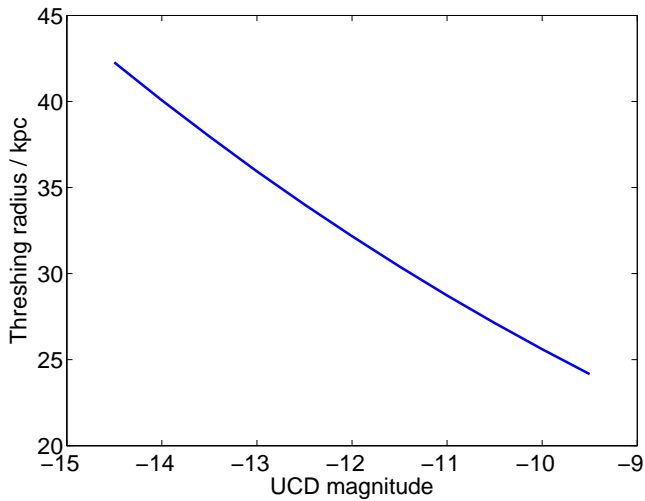,width=8.6cm}
\caption{The predicted threshing radius as a function of UCD
  B-magnitude.}
\label{fig:threshrad}
\end{figure}

This fixes the mean density within the core using
equation~\ref{eq:rhodmobs}, and we insert this in
equation~\ref{eq:tidal} to determine the threshing radius of any UCD
as a function of its absolute magnitude.  The results of these
calculations are shown in Fig.~\ref{fig:threshrad}.

Our approach in estimating the threshing radii is very similar to that
used by \citet{BCD03} except for the following differences. First, we
specifically use the local gradient of the cluster potential (rather
than the point-mass assumption) and we add a core component to the
cluster NFW potential. Secondly, we have used more recent scaling
relations to derive the dwarf galaxy core masses as a function of
their absolute magnitudes. Our estimates give very similar threshing
radii: compare our Fig.~\ref{fig:threshrad} with their fig.~7.

Recently, \citet[hereafter GMK07]{GMK07} have conducted
numerical simulations of threshing in a static potential similar to
that of the Virgo Cluster.  They use two different models of a dwarf
galaxy: one which consists solely of an extended dark matter halo with
an NFW profile, and one in which this profile has been
centrally-concentrated by a dissipative baryonic disk.  For the
latter, the threshing radii they find are similar to ours.  The
dark-matter-only halos can have much greater threshing radii, up to
200\,kpc, but only for galaxies on quite circular orbits.  As we
discuss in Section~\ref{sec:conc} below, the two models bracket our
predictions for UCD fractions as a function of radius, and both lead
to the same qualitative results.

The detailed threshing simulations of individual dwarf galaxies by
\citet{BCD03} showed that several pericentre passages within the
threshing radius were necessary to completely strip the dwarf galaxy.
In our model below we do not count the number of orbits, but simply
assume that any galaxy with an orbit that passes within its threshing
radius will be stripped.  This assumption is reasonable for
galaxies within about 100\,kpc of the cluster centre, but for those
galaxies with radii of order 300\,kpc, on the outskirts of the
observed UCD distribution, there may have been only a single
pericentric passage in the lifetime of the cluster.  This could lead to
an over-estimate of the UCD fraction at large radii and would
strengthen our results.

\subsection{Galaxy orbits}
\label{sec:orbits}

The equation of motion for galaxy orbits in a spherically-symmetric
potential is:
\begin{equation}
\ddot{r}-{L^2\over r^3}=-{GM_r\over r^2}
\label{eq:orbit}
\end{equation}
where $L=rv_t=$\,const is the specific angular momentum and $v_t$ is the
tangential component of the velocity.

Combining equations~\ref{eq:jeans} and \ref{eq:orbit}, multiplying by
$\dot{r}$ and integrating leads to the following energy equation:
\begin{equation}
{1\over2}\dot{r}^2+{1\over2}{L^2\over r^2}-\sigma_r^2\,\ln(\rho\sigma_r^2)
-2\beta\,\sigma_r^2\,\ln r={\rm const},
\end{equation}
where we have taken $\beta$ to be constant and used the approximation
that the gradient in $\sigma_r^2$ is much less than that in $\rho$ and
can be neglected.
Putting in initial conditions (labelled with subscript
0) and setting $\dot{r}=0$, the following equation is obtained for
the minimum and maximum values of $r$:
\begin{equation}
\left(r_0\over r\right)^2=\left(v_0\over v_{t0}\right)^2
+2\left(\sigma_r\over v_{t0}\right)^2\,
\left[\ln\left(\rho\over\rho_0\right)
+\beta\,\ln\left(r\over r_0\right)^2\right].
\end{equation}
Simple iteration of this equation quickly finds the minimum orbital
radius (pericentre).

\begin{figure}
\psfig{file=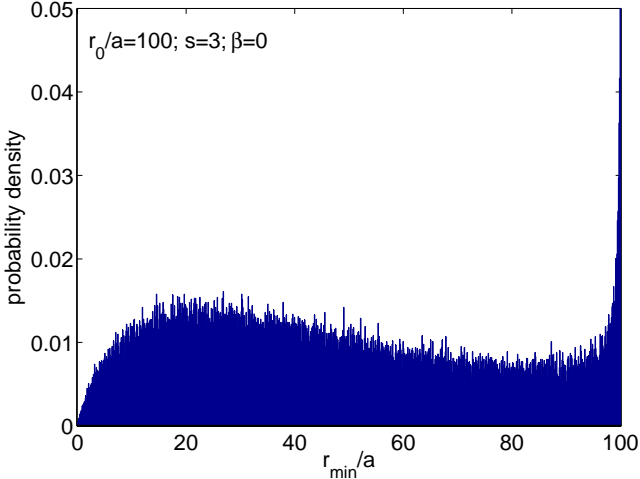,width=8.6cm}
\caption{A histogram showing the probability density  for the
  distribution of minimum orbital radii (pericentres) for a selection
  of galaxy orbits drawn from the appropriate Gaussian distribution of
  velocities.  For this particular example the 1 percentile of minimum
  radii is at $r_{\rm min}\approx0.03\,r_0$.}
\label{fig:minrad}
\end{figure}

For each value of $r_0$, we draw 10\,000 velocities with the
appropriate Gaussian distributions in each of the radial and
tangential directions, then solve for the pericentric radius.  An
example histogram is shown in Fig.~\ref{fig:minrad}.  The sharp
spike at $r_{min}=r_0$ is because any orbit that has
an initial radial velocity close to zero and a tangential velocity
that exceeds the circular velocity at that radius will already be at
pericentre.  More importantly, there is a wide distribution of minimum
radii extending all the way down to $r_{\rm min}\approx0$, even for an
isotropic velocity dispersion tensor.  This orbital 
distribution is in good agreement with that found in cosmological 
simulations, for example by \citet{GMG98}.

\begin{figure}
\psfig{file=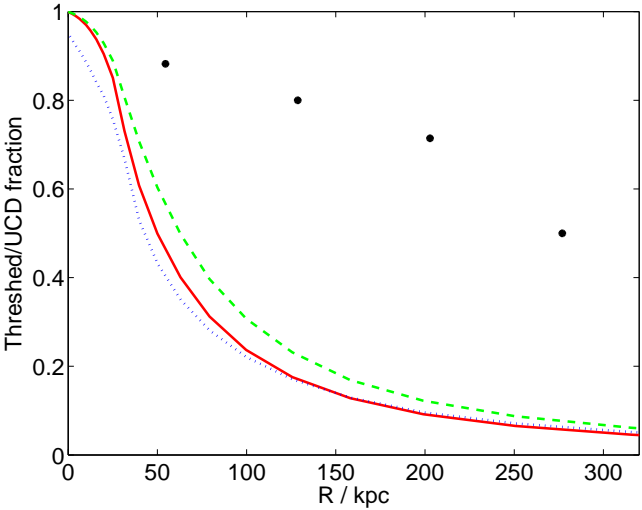,width=8.6cm}
\caption{Comparison of observed and predicted fractions of UCDs.  The
circles show the observed UCD fraction (of the joint UCD plus bright
dwarf sample) as a function of projected radius from the centre of the
cluster.  The lines show the predicted fraction of orbits that pass
within the threshing radius of 30\,kpc for $s=3$, $a=5$\,kpc,
$\beta=0$, (solid, red), $s=$, $a=5$\,kpc, $\beta=0.5$, (dashed,
green), and $s=4$, $a=90$\,kpc, $\beta=0.$, (dotted, blue).}
\label{fig:fraction_2d_compare}
\end{figure}

The variation in threshing radii for different galaxies is small and
so for simplicity we adopt a constant value of 30\,kpc.  Then for each
radius, $r$, we can tabulate the fraction of orbits that pass within
this radius.  This can then be projected along the line-of-sight with
the appropriate density weighting to determine the fraction of
threshed orbits as a function of projected distance from the cluster
centre.  The results of this calculation are shown in
Fig.~\ref{fig:fraction_2d_compare} for the two example density
profiles discussed above.

It is immediately apparent that the predicted fraction of threshed
galaxies is far too low at radii greater than about 50\,kpc.  The
predicted UCD fraction drops rapidly at this radius, whereas the
observed fraction of UCDs stays high out to 250\,kpc.  (We have checked
that this conclusion is unaltered even if the UCDs are distributed on
a plane perpendicular to the line-of-sight such that their projected
radii are equal to the true distances from the centre of the cluster.)
There are many simplifications and uncertainties in the model, but it
is hard to see how these could make a difference of a factor of five.
The static threshing model is simply untenable.

\section{Summary and Conclusions}
\label{sec:conc}

In this paper we have investigate the possibility that UCD galaxies
are formed by the threshing of nucleated, early-type dwarf galaxies.

We first contrast the distribution of nucleated and non-nucleated
dwarfs, which are indistinguishable apart from a small excess of
bright, nucleated dwarfs at small clustocentric radii.  We concur with
the conclusion of \citet{CPF06} that the observations are consistent
with a single population in which all dwarfs are nucleated, with a
ratio of nuclear to total magnitude that varies slowly with
magnitude.  However, we need to flatten their relation in order to
obtain a good fit when extrapolating to fainter magnitudes.

Given this hypothesis, we can reproduce the magnitude distribution of
the UCD population, except at bright magnitudes where the model
predicts more UCDs than are observed.  Under the threshing model, the
UCDs are likely to have originated from dwarfs with magnitudes brighter
than about $M_B=-15$.  We use the joint UCD plus bright dwarf
population in the modelling that follows.

The threshing model predicts that over half of all dwarf galaxies must
be disrupted: 38 surviving dwarfs have nuclei of similar magnitude to
the 49 observed UCDs.  This may seem excessive but corresponds to an
intracluster light fraction of just 8 per cent, well within the
observed range for clusters of this mass \citep{FCJ04,FMM04,GZZ05,Zib05}.

The distribution of dwarf galaxies in Fornax follows that of the total
mass distribution and shows no evidence for disruption of dwarfs near
the cluster core.  Nevertheless, the UCD population is more
centrally-concentrated than the dwarfs, as would be expected in the
threshing model.  If we assume that the joint population is in a steady-state
dynamically, then it should also satisfy the Jeans Equations.  We show
that the joint population is well-described by a density distribution
of the form
\begin{equation}
\rho=\frac{\rho_0}{x\,(1+x)^{s-1}},
\end{equation}
where $x=r/a$, and $a$ and $s$ are constants, with $s$ lying between
about 3 and 4.5.

The velocity dispersion of UCD galaxies shows a sharp decline with
radius that is hard to explain.  It may in part be due to a radial
bias in the orbits, but this is not enough in itself to explain the
effect.  The velocity dispersion of bright dwarfs is greater than that
of the UCDs.  When the two are combined, then the joint population
with density slope $s=3$ provides a marginally-acceptable fit to the mass
profile of Fornax.

We have tested the simplest possible threshing model, in which dwarf
galaxies move on orbits in a static cluster potential and are threshed
if they pass within a radius at which the tidal force from the cluster
exceeds the internal gravity at the core their dark matter halo.  This
fails to reproduce the observed fraction of UCDs at radii greater than
50\,kpc from the core of Fornax.  There are many deficiencies in the
model but these are unlikely to raise the threshing radii by a factor
of 5, as is required, and so we conclude that this static mode is unviable.

Our results have several points of agreement with the earlier work by
\citet{BCD03} despite a very different approach: we have used analytic
descriptions of the cluster dynamics compared to their numerical
computations.  In our work we have based our prediction on a parent
sample of dwarf galaxies generated directly from the known Fornax
galaxies, whereas Bekki et al.\ generated their galaxy sample from
more general empirical relations for the luminosity functions and
radial profiles of galaxies within clusters.  In particular, they used
a King profile with a core radius of 50\,kpc for the density
distribution, very different from our model.  They demonstrate that
dwarf galaxies are disrupted if they pass inside their critical
threshing radius when orbiting the cluster centre. They then use this
radius to estimate the population of threshed galaxies (UCDs) in the
Fornax Cluster.  They find this to be consistent with the known
distribution of the 7 very luminous UCDs known in the cluster at that
time.

Our conclusion (refuting the simple threshing model) differs from that
of Bekki et al.\ for a number of reasons.  Firstly, we use the
measured positions of galaxies in Fornax, rather than a generic King
model.  We also have many fewer dwarf galaxies than predicted by their
Schechter model for the cluster luminosity function.  In addition, we
have extended the analysis to much lower luminosities of both the UCDs
(as new data have become available) and the parent galaxies (due to
the greater difference in nuclear to total luminosity now used).  This
new analysis has clearly revealed a disagreement between the number of
UCDs at large clustocentric distances and the threshing predictions.

A recent paper by GMK07 undertook an extensive series of simulations
to investigate the disruption of UCD host galaxies within a cluster
potential similar to that of the Virgo cluster.  They considered two
different models for the host galaxy with very different degrees of
central concentration and followed their threshing in a static
potential over 5 Gyr.  They then looked at the orbits of particles in
a cosmological simulation of cluster formation to assess which of
those orbits would lead to threshing.  This latter step follows the
dynamical evolution of the halo and is much more realistic than a
static potential.

GMK07 state that their model ``leads to the observed spatial
distribution of UCDs'', in apparent disagreement with our results
above.  In fact our theoretical UCD fractions as a function of radius
agree with theirs and are bracketed by their upper and lower
predictions.  The difference in the conclusion arises from the very
different observed threshing fractions that we adopt.  GMK07 use only
15 UCDs in both Virgo and Fornax combined, whereas we use a new sample
of 49 UCDs from Fornax alone.  Also, GMK07 do not say how they define
the nucleated dwarfs corresponding to the parent sample, whereas we
are careful to select only those dwarfs that would have nuclei that
match those of the observed UCDs.

In conclusion, the origin of UCDs as dwarf galaxy nuclei remains
unproven.  Our modelling has revealed a number of attractive features:
\begin{description}
\item The distribution of nuclear magnitudes for dwarf galaxies
  roughly matches that of known UCD galaxies.
\item UCDs are more centrally-concentrated within Fornax than are
  dwarf galaxies.  (However, this would also be true if the UCDs
  constituted an extended globular cluster population around
  NGC\,1399.)
\item The joint UCD plus bright dwarf population has a smooth density
profile with a recognisable (NFW) form and appears to sit in dynamical
equilibrium within the Fornax cluster.
\end{description}
At the same time, there are several major deficiencies in the model:
\begin{description}
\item The model requires that more dwarf galaxies must have been
  disrupted in Fornax than currently remain.  However, the spatial
  distribution of dwarfs matches that of the total mass profile of the
  cluster and shows no sign of galaxy disruption near the cluster
  core.
\item The very low velocity dispersion of UCDs as compared to bright
  dwarfs is unexplained, as is the sharp decline in velocity
  dispersion of the UCDs with radius.  (However, this would prove true
  for any dynamical model of the UCD population, regardless of its
  origin.)
\item A static threshing model for UCD formation, based upon orbits
  within the current cluster potential, is a hopeless failure.  It
  predicts far too few UCDs at radii greater than about 30\,kpc.
\item The simulations of GMK07 within an evolving cluster
  potential also give too few UCDs at large radii.
\end{description}
The balance of evidence would seem to be against the threshing model.
Before dismissing the model altogether, however, we note that the
threshing may have occurred within smaller sub-clusters that later
fell into Fornax and have not yet reached dynamical equilibrium.
This mechanism is suggested by the spatial distribution of UCDs in the
Fornax Cluster: they show some association with normal galaxies and,
in particular, lie in a band across the cluster \citep{Gre08}.

In considering the threshing hypothesis for UCD formation we should
not discuss the dynamical properties of the objects in isolation from
their internal properties.  \citet{EGD07} studied the stellar
populations of Virgo Cluster UCDs and concluded that the Virgo UCDs
have stellar populations the globular clusters of the central galaxies
M87 and M49 (old ages, a range of metallicity, and supersolar
alpha-abundances).  On this basis, the Virgo UCD stellar populations
are not consistent with simple threshing model.  On the other hand,
\citet{MHI06} found metallicities and (a range of) ages in Fornax
Cluster UCDs, which are more in agreement with the hypothesis that the
Fornax UCDs are threshed nuclei.  A detailed analysis of the structure
and colours of both Virgo and Fornax UCDs \citep{Evs08} concluded that
their structural properties could be consistent with either globular
clusters or dwarf galaxy nuclei, with the interesting observation that
UCDs are about twice as extended (in effective radius) as the nuclei
of dwarf galaxies at the same luminosity.

Most of these observational results, as well as our own analysis in this
paper, argue against the simple threshing hypothesis for UCD formation.

\section*{Acknowledgements}

This work was initated while PAT was a visitor to Queensland under ARC
Discovery Projects Grant DP0557676, and continued whist MJD was
visiting Sussex with partial funding from PPARC Grant PP/D001579/1.
We would like to thank an anonymous referee for many useful
suggestions for comparison with previous work.

\appendix
\section{Tables of galaxy and UCD data}
\label{sec:datatables}


\begin{table*}\begin{minipage}{160mm}
\caption{The galaxy sample.}

\begin{tabular}{rrrrrrrrrrrl}
 FCC& X& Y& R& cz& err& ref& $M_B$& nuc& late& dwarf& morph\\
 & kpc& kpc& kpc& km/s& km/s& & mag& & & & \\
\\
     1&  -1968.9&    709.2&   2070.2&     - &     - 
&   - &    -12.8&   1&    0&    1& dE,N\\
     2&  -1656.6&    666.0&   1767.9&     4540&        9
&      1&    -16.2&   0&    1&    1& d:SBc(LSB)\\
     3&  -1532.9&    -32.6&   1534.0&     1567&        1
&      1&    -17.2&   0&    1&    0& SBcdII-III\\
     4&  -1446.4&   -593.1&   1577.8&     - &     - 
&   - &    -14.3&   0&    0&    1& dE2\\
     5&  -1441.4&   -459.9&   1524.4&     - &     - 
&   - &    -14.0&   0&    0&    1& dE5/ImV\\
     6&  -1460.3&    -70.7&   1463.6&     - &     - 
&   - &    -15.3&   0&    0&    1& dE5\\
     7&  -1399.5&    -33.7&   1400.6&     - &     - 
&   - &    -12.4&   0&    0&    1& dE1\\
     8&  -1427.5&    741.2&   1592.6&     - &     - 
&   - &    -12.8&   0&    0&    1& dE?\\
     9&  -1400.7&    976.0&   1688.6&     1751&        3
&      1&    -16.0&   0&    1&    0& Sd?\\
    10&  -1392.4&   1041.2&   1719.4&     1451&        3
&      1&    -16.8&   0&    1&    0& Sd(onedge)\\
    11&  -1297.6&    306.7&   1326.5&     - &     - 
&   - &    -14.3&   1&    0&    1& dE4,N\\
    12&  -1292.1&    752.2&   1480.9&     - &     - 
&   - &    -14.3&   0&    1&    0& BCDorS\\
    13&  -1215.2&   -589.4&   1362.5&     1792&       25
&      2&    -19.3&   0&    1&    0& SBcII\\
    14&  -1215.9&   -285.3&   1254.9&     1805&       10
&      1&    -13.7&   0&    0&    1& dE3\\
    15&  -1224.6&    696.6&   1396.3&     - &     - 
&   - &    -14.7&   0&    0&    1& dE7\\
    16&  -1184.2&   -166.2&   1199.2&     - &     - 
&   - &    -13.5&   0&    0&    1& dE2\\
    17&  -1176.6&    604.5&   1312.0&     - &     - 
&   - &    -13.3&   1&    0&    1& dE3,N\\
    18&  -1156.8&    112.9&   1159.9&     - &     - 
&   - &    -15.8&   0&    1&    0& SmIII\\
    19&  -1116.3&   -679.7&   1319.0&     1497&       47
&      2&    -16.1&   1&    0&    1& dS0(8),N\\
    20&  -1111.8&   -600.5&   1274.4&     - &     - 
&   - &    -12.8&   0&    0&    1& filament\\
    21&  -1096.7&   -614.3&   1268.0&     1751&       13
&      3&    -21.9&   0&    0&    0& S0(pec)\\
    22&  -1095.5&   -577.5&   1248.7&     1979&       11
&      3&    -19.4&   0&    1&    0& Sapec\\
    23&  -1051.9&   -501.7&   1174.2&     - &     - 
&   - &    -12.7&   1&    0&    1& ImVordE5,N\\
    25&  -1040.3&   -534.9&   1178.8&     - &     - 
&   - &    -13.6&   1&    0&    1& dE0,N\\
    26&  -1052.2&   -114.7&   1060.5&     1786&       28
&      3&    -16.3&   0&    0&    0& SB0(8)\\
    27&  -1042.7&    223.9&   1062.4&     - &     - 
&   - &    -12.0&   0&    0&    1& dE2\\
    28&  -1009.1&   -718.8&   1250.0&     1408&       23
&      3&    -17.7&   0&    1&    0& SmIII\\
    29&  -1020.8&   -354.5&   1086.8&     1368&       26
&      3&    -19.5&   0&    1&    0& SBa(r)\\
    30&  -1004.3&   -619.8&   1190.0&     - &     - 
&   - &    -12.4&   0&    0&    1& dEordS0\\
    31&  -1010.8&     65.4&   1011.8&     1542&       42
&      3&    -14.5&   0&    0&    1& dE4\\
    32&   -967.6&      5.2&    967.5&     1342&       27
&      3&    -15.9&   0&    0&    1& dEpec/BCD\\
    33&   -941.4&   -544.3&   1095.6&     1990&       14
&      3&    -17.1&   0&    1&    0& SdIIIpec/BCD\\
    34&   -958.4&     78.7&    960.3&     - &     - 
&   - &    -12.9&   0&    0&    1& dE\\
    35&   -935.8&   -515.7&   1076.2&     1841&       22
&      3&    -16.0&   0&    1&    0& SmIV/BCD?\\
    36&   -973.0&    889.3&   1307.5&     - &     - 
&   - &    -15.4&   1&    0&    1& dE4pec,N\\
    37&   -936.9&   -319.4&    994.9&     1924&       65
&      3&    -17.5&   0&    1&    0& SBcIII(interacting)\\
    38&   -934.1&   -385.0&   1016.4&     - &     - 
&   - &    -13.8&   0&    0&    1& dE0pec\\
    39&   -924.1&   -326.2&    985.1&     1007&       18
&      3&    -15.8&   0&    1&    0& SdIII(interacting)\\
    40&   -913.8&   -451.8&   1026.2&     - &     - 
&   - &    -13.7&   0&    0&    1& dE4\\
    41&   -943.4&    869.1&   1272.7&     - &     - 
&   - &    -13.7&   0&    1&    1& ImVordE3\\
    42&   -903.0&    -20.7&    903.5&     - &     - 
&   - &    -13.0&   0&    0&    1& dE0\\
    43&   -911.7&    892.0&   1265.9&     1323&       17
&      2&    -17.8&   1&    0&    1& dS0/2(5),N\\
    44&   -882.0&    112.2&    887.3&     1232&       32
&      3&    -13.8&   0&    0&    1& dS0\\
    45&   -879.7&    312.9&    929.1&     - &     - 
&   - &    -11.8&   0&    1&    1& ImV\\
    46&   -839.4&   -585.5&   1030.9&     2255&       27
&      3&    -15.7&   0&    0&    1& dE4\\
    47&   -846.6&    -91.9&    852.9&     1434&       10
&      3&    -18.0&   0&    0&    0& E4\\
    48&   -846.1&    314.8&    898.4&     1439&       45
&      3&    -14.2&   0&    0&    1& dE3\\
    49&   -816.1&   -645.3&   1048.1&     - &     - 
&   - &    -12.1&   0&    0&    1& dE4\\
    50&   -822.6&    -23.8&    823.2&     - &     - 
&   - &    -14.7&   1&    0&    1& dE0,N\\
    51&   -810.3&   -470.7&    943.2&     - &     - 
&   - &    -12.8&   0&    0&    1& dE4\\
    52&   -817.1&    366.7&    890.9&     - &     - 
&   - &    -12.8&   0&    0&    1& dE1\\
    53&   -816.0&    685.8&   1058.6&     1675&       35
&      3&    -16.6&   0&    1&    0& ScdIII\\
    54&   -819.1&    842.9&   1167.3&     - &     - 
&   - &    -13.3&   1&    0&    1& dE1,N\\
    55&   -804.0&    322.6&    862.1&     1252&       10
&      3&    -17.4&   1&    0&    0& S0(9),N\\
    56&   -783.9&   -243.0&    823.9&     - &     - 
&   - &    -13.8&   1&    0&    1& dE1,N\\
    57&   -778.7&   -165.2&    798.2&     - &     - 
&   - &    -13.5&   0&    0&    1& dE6(boxy)pec\\
    58&   -754.9&   -717.1&   1048.6&     - &     - 
&   - &    -13.3&   0&    0&    1& dE2\\
    59&   -777.3&    658.1&   1011.9&     - &     - 
&   - &    -11.9&   1&    0&    1& dE0,N\\
    60&   -733.8&   -672.1&   1001.9&     - &     - 
&   - &    -12.9&   0&    0&    1& ImVordE2\\
    61&   -761.3&    623.6&    977.8&     - &     - 
\end{tabular}
\end{minipage}\end{table*}
  
\begin{table*}\begin{minipage}{160mm}
\contcaption{}

\begin{tabular}{rrrrrrrrrrrl}
 FCC& X& Y& R& cz& err& ref& $M_B$& nuc& late& dwarf& morph\\
 & kpc& kpc& kpc& km/s& km/s& & mag& & & & \\
\\
&   - &    -13.3&   1&    0&    1& dE3,N\\
    62&   -731.2&   -593.1&    947.8&     1878&       30
&      3&    -18.7&   0&    1&    0& SbcII\\
    63&   -765.2&   1104.5&   1335.8&     1354&       19
&      1&    -18.6&   0&    0&    0& E4\\
    64&   -715.7&  -1078.8&   1302.8&     - &     - 
&   - &    -13.8&   0&    0&    1& dE5\\
    65&   -739.3&     74.6&    742.1&     - &     - 
&   - &    -13.8&   1&    0&    1& dE6,NordS0\\
    66&   -726.8&    599.0&    936.1&     - &     - 
&   - &    -14.4&   0&    0&    1& dE3\\
    67&   -689.8&     94.3&    695.0&     1400&       21
&      3&    -18.3&   0&    1&    0& Sc(onedge)\\
    68&   -697.1&    560.0&    889.0&     2030&       26
&      3&    -14.8&   1&    0&    1& dE5,N\\
    69&   -669.4&   -344.6&    756.6&     - &     - 
&   - &    -12.6&   0&    0&    1& ImVordE0\\
    70&   -655.7&   -666.9&    940.9&     - &     - 
&   - &    -12.8&   1&    0&    1& dE2,N?\\
    71&   -641.8&    -53.5&    644.6&     - &     - 
&   - &    -10.8&   0&    0&    1& dE2\\
    72&   -631.3&   -145.7&    649.5&     - &     - 
&   - &    -12.0&   0&    0&    1& dE0/ImV\\
    73&   -623.3&   -437.3&    765.5&     - &     - 
&   - &    -11.9&   0&    0&    1& dE3\\
    74&   -650.4&   1217.1&   1374.0&     - &     - 
&   - &    -13.6&   0&    0&    1& dE\\
    75&   -629.6&     17.6&    629.6&     - &     - 
&   - &    -11.5&   0&    0&    1& dE3\\
    76&   -637.1&    660.8&    913.0&     1868&       42
&      3&    -16.6&   0&    1&    1& ImII/dS0(6)emission\\
    77&   -618.4&   -269.2&    677.2&     - &     - 
&   - &    -13.1&   1&    0&    1& dE0,N\\
    78&   -620.8&     25.1&    621.1&     - &     - 
&   - &    -12.1&   0&    0&    1& ImVordE\\
    79&   -616.1&   -211.6&    653.7&     - &     - 
&   - &    -12.6&   0&    0&    1& dE?\\
    80&   -614.7&   1038.6&   1201.5&     - &     - 
&   - &    -15.7&   0&    0&    1& dEordS0\\
    81&   -603.7&    613.3&    856.2&     1893&       36
&      3&    -14.2&   1&    0&    1& dE1,N\\
    82&   -575.2&    415.5&    706.2&     1157&       50
&      4&    -14.9&   1&    0&    1& dE1,N\\
    83&   -565.5&    207.7&    600.5&     1431&       11
&      3&    -19.0&   0&    0&    0& E5\\
    84&   -562.5&    142.7&    578.9&     - &     - 
&   - &    -11.8&   1&    0&    1& dE0,N?\\
    85&   -547.6&    -34.6&    549.1&     1673&       69
&      4&    -15.0&   1&    0&    1& dE0,N\\
    86&   -548.4&     33.5&    549.1&     - &     - 
&   - &    -13.8&   1&    0&    1& dE5,N?\\
    87&   -548.6&    660.9&    855.0&     - &     - 
&   - &    -12.7&   0&    0&    1& dE3\\
    88&   -534.0&    635.5&    826.3&     1829&       13
&      3&    -19.5&   0&    1&    0& SBb(r)I\\
    89&   -531.2&    770.3&    931.8&     - &     - 
&   - &    -13.3&   1&    0&    1& dE,N\\
    90&   -516.8&   -293.2&    596.5&     1813&       15
&      2&    -16.3&   0&    0&    0& E4pec\\
    91&   -529.9&   1185.8&   1294.6&     1590&       12
&      1&    -14.3&   0&    1&    1& ImV\\
    92&   -515.1&    171.0&    541.3&     - &     - 
&   - &    -12.3&   0&    0&    1& ImVordE\\
    93&   -505.1&   -128.9&    522.4&     - &     - 
&   - &    -11.9&   0&    0&    1& dE2\\
    94&   -507.9&    167.1&    533.3&     - &     - 
&   - &    -12.0&   0&    0&    1& dE0\\
    95&   -503.3&     41.7&    504.7&     1276&       12
&      3&    -16.7&   0&    0&    1& dSB0ordSBa\\
    96&   -491.3&   -754.7&    904.4&     - &     - 
&   - &    -12.1&   0&    0&    1& dE0/ImV\\
    97&   -499.9&    -15.4&    500.2&     - &     - 
&   - &    -12.3&   0&    0&    1& dE1\\
    98&   -480.6&   -288.4&    562.6&     - &     - 
&   - &    -13.1&   0&    0&    1& ImVordE1\\
    99&   -485.3&    388.2&    618.9&     - &     - 
&   - &    -14.2&   0&    0&    1& dE4\\
   100&   -477.9&    139.2&    496.6&     1660&       31
&      2&    -15.8&   1&    0&    1& dE4,N\\
   101&   -475.1&    -78.7&    482.2&     1051&       55
&      3&    -14.1&   1&    0&    1& dE0,NorS\\
   102&   -443.8&   -268.9&    520.8&     1723&       61
&      2&    -14.8&   0&    1&    1& ImIV\\
   103&   -426.8&   -113.5&    442.4&     - &     - 
&   - &    -12.0&   0&    0&    1& dE2\\
   104&   -430.6&    384.8&    575.4&     - &     - 
&   - &    -12.8&   0&    0&    1& dE6\\
   105&   -422.0&   -222.7&    478.6&     - &     - 
&   - &    -14.0&   1&    0&    1& dE1,N?\\
   106&   -410.3&    423.0&    587.2&     2066&       11
&      3&    -16.2&   1&    0&    1& d:S0(6),N\\
   107&   -395.0&   -832.9&    924.5&     - &     - 
&   - &    -14.3&   1&    0&    1& dE3,N?\\
   108&   -399.6&   -245.3&    470.4&     - &     - 
&   - &    -12.2&   0&    0&    1& dE0\\
   109&   -394.0&   -765.0&    863.1&     - &     - 
&   - &    -13.0&   0&    0&    1& dE4\\
   110&   -391.6&   -100.3&    404.9&     - &     - 
&   - &    -14.5&   0&    0&    1& dE4\\
   111&   -393.6&    602.5&    717.4&     1283&      115
&      4&    -14.5&   1&    0&    1& dE0,N\\
   112&   -381.0&   -346.6&    516.9&     - &     - 
&   - &    -14.2&   1&    0&    1& dS0(5),N\\
   113&   -384.7&    223.9&    443.9&     1416&       20
&      3&    -16.1&   0&    1&    0& ScdIIIpec\\
   114&   -379.7&     18.6&    380.1&     - &     - 
&   - &    -11.6&   0&    0&    1& dE1?\\
   115&   -377.6&    -93.5&    389.6&     1686&       25
&      3&    -14.7&   0&    1&    0& Sdm(onedge)\\
   116&   -371.9&   -197.9&    422.5&     1204&       57
&      3&    -15.2&   1&    0&    1& dE1,N\\
   117&   -361.0&   -827.6&    905.2&     - &     - 
&   - &    -13.1&   0&    0&    1& dE0\\
   118&   -357.1&    347.3&    496.6&     - &     - 
&   - &    -13.7&   1&    0&    1& dE0,N\\
   119&   -357.6&    655.3&    744.7&     1417&       18
&      3&    -16.3&   0&    0&    0& S0pec\\
   120&   -344.1&   -403.2&    531.8&      887&        6
&      1&    -15.0&   0&    1&    1& ImIV\\
\end{tabular}
\end{minipage}\end{table*}
  
\begin{table*}\begin{minipage}{160mm}
\contcaption{}

\begin{tabular}{rrrrrrrrrrrl}
 FCC& X& Y& R& cz& err& ref& $M_B$& nuc& late& dwarf& morph\\
 & kpc& kpc& kpc& km/s& km/s& & mag& & & & \\
\\
   121&   -343.8&   -241.2&    421.2&     1446&       12
&      2&    -21.1&   0&    1&    0& SBbc(s)I\\
   122&   -351.8&   1056.1&   1111.3&     - &     - 
&   - &    -14.0&   0&    0&    1& dS0\\
   123&   -336.9&   -143.0&    366.8&      940&       21
&      3&    -13.4&   0&    1&    1& ImV\\
   124&   -336.9&    447.1&    558.2&     - &     - 
&   - &    -13.3&   0&    0&    1& dE3+ImV\\
   125&   -330.7&   -134.6&    357.8&     - &     - 
&   - &    -12.6&   0&    0&    1& dE4\\
   126&   -328.6&    385.3&    504.9&     - &     - 
&   - &    -10.9&   0&    0&    1& dE0\\
   127&   -312.3&     60.6&    317.8&     - &     - 
&   - &    -11.5&   0&    0&    1& dE3?\\
   128&   -306.4&   -354.5&    469.8&     - &     - 
&   - &    -14.5&   0&    1&    1& ImIV\\
   130&   -307.5&    -23.2&    308.5&     - &     - 
&   - &    -13.1&   0&    1&    1& ImV?\\
   131&   -305.1&     77.6&    314.3&     - &     - 
&   - &    -11.0&   0&    0&    1& dE3\\
   132&   -295.7&   -120.2&    319.8&     1883&       98
&      4&    -12.8&   0&    0&    1& dE2\\
   133&   -295.1&     30.8&    296.5&     - &     - 
&   - &    -13.8&   1&    0&    1& dE0,N\\
   134&   -296.0&    299.4&    419.9&     1381&       19
&      3&    -13.7&   1&    0&    1& dE5pec,NorE\\
   135&   -286.1&    402.4&    492.6&     1232&       18
&      3&    -15.8&   1&    0&    1& dS0(5),N\\
   136&   -283.4&    -33.5&    285.5&     1206&       23
&      3&    -16.5&   1&    0&    1& dE2,N\\
   137&   -265.0&   -143.5&    302.0&     - &     - 
&   - &    -14.4&   1&    0&    1& dE0,N\\
   138&   -254.9&   -304.7&    398.2&     - &     - 
&   - &    -12.7&   0&    0&    1& dE2\\
   139&   -259.2&    981.1&   1013.8&     1752&       24
&      2&    -16.9&   0&    1&    0& SBmIII\\
   140&   -252.6&     90.5&    267.9&     - &     - 
&   - &    -12.3&   1&    0&    1& dE4,N\\
   142&   -250.9&    142.4&    287.9&     - &     - 
&   - &    -12.8&   1&    0&    1& dE2,N\\
   143&   -249.4&     97.6&    267.5&     1356&       10
&      3&    -17.0&   0&    0&    0& E3\\
   144&   -247.7&     44.7&    251.5&     - &     - 
&   - &    -12.1&   0&    0&    1& dE0\\
   145&   -241.7&     81.0&    254.7&     - &     - 
&   - &    -11.7&   0&    0&    1& dE0\\
   146&   -234.3&     44.4&    238.3&     - &     - 
&   - &    -11.8&   1&    0&    1& dE4,N\\
   147&   -228.2&     77.8&    240.8&     1340&       12
&      3&    -19.4&   0&    0&    0& E0\\
   148&   -228.1&     64.1&    236.6&      749&       10
&      3&    -17.7&   0&    0&    0& S0(cross)\\
   149&   -217.6&   -223.8&    312.8&     - &     - 
&   - &    -11.4&   0&    0&    1& dE\\
   150&   -216.6&   -318.8&    386.1&     1411&       18
&      3&    -15.6&   1&    0&    1& dE4,N\\
   151&   -215.5&   -254.2&    334.0&     - &     - 
&   - &    -13.3&   1&    0&    1& dE0,N?\\
   152&   -215.7&   1042.5&   1063.9&     1389&       12
&      2&    -17.2&   0&    1&    0& S0/apec;emission\\
   153&   -213.4&    350.2&    409.4&     1589&       10
&      3&    -18.3&   0&    0&    0& S0(9)\\
   154&   -212.0&     69.5&    222.8&     - &     - 
&   - &    -12.1&   0&    0&    1& dE3\\
   155&   -209.0&    225.4&    306.8&     - &     - 
&   - &    -13.0&   0&    0&    1& dE2\\
   156&   -197.2&     39.2&    200.9&     - &     - 
&   - &    -14.1&   0&    0&    1& dE1\\
   157&   -196.7&    -22.2&    198.0&     - &     - 
&   - &    -13.0&   1&    0&    1& dE0,N?\\
   158&   -191.4&   -188.3&    268.9&     - &     - 
&   - &    -14.5&   1&    0&    1& dE6,N\\
   159&   -183.0&    217.3&    283.7&     - &     - 
&   - &    -13.2&   1&    0&    1& dE3,N\\
   160&   -171.9&     21.5&    173.1&     - &     - 
&   - &    -13.6&   1&    0&    1& dE1,N\\
   161&   -171.7&      2.6&    171.7&     1351&       10
&      3&    -19.6&   0&    0&    0& E0\\
   162&   -167.8&      5.9&    167.9&     - &     - 
&   - &    -11.3&   0&    0&    1& dE0\\
   163&   -167.5&   -137.0&    216.7&     - &     - 
&   - &    -11.5&   0&    0&    1& dE0\\
   164&   -159.8&   -250.0&    297.1&     1427&       49
&      3&    -14.9&   1&    0&    1& dS0(5),N\\
   165&   -147.7&   -160.9&    218.8&     - &     - 
&   - &    -13.8&   0&    0&    1& dE6\\
   166&   -144.9&   -598.1&    615.8&     - &     - 
&   - &    -13.2&   1&    0&    1& dE3,N\\
   167&   -144.6&    165.3&    219.4&     1953&       13
&      2&    -20.0&   0&    0&    0& S0/a\\
   168&   -143.9&     83.7&    166.3&     - &     - 
&   - &    -12.6&   0&    0&    1& dE0\\
   169&   -140.6&    226.2&    266.1&     - &     - 
&   - &    -14.1&   0&    0&    1& dE2\\
   170&   -139.2&     53.6&    149.1&     1744&       10
&      3&    -18.3&   0&    0&    0& S0(9)(boxy)\\
   171&   -132.2&     22.5&    134.0&     - &     - 
&   - &    -13.4&   1&    0&    1& dE0,N?\\
   172&   -129.8&   -668.4&    681.1&     - &     - 
&   - &    -10.9&   0&    0&    1& dE4\\
   173&   -127.4&    450.7&    468.1&     - &     - 
&   - &    -13.3&   0&    0&    1& dE3/ImV\\
   174&   -126.3&    850.6&    859.7&     1801&       46
&      3&    -14.6&   1&    0&    1& dE1,NorE(cD)\\
   175&   -125.8&      5.2&    125.9&     - &     - 
&   - &    -12.5&   0&    0&    1& dE3\\
   176&   -121.8&   -281.3&    306.8&     1410&       10
&      3&    -17.6&   0&    1&    0& SBa(s)\\
   177&   -121.3&    248.3&    276.1&     1558&        9
&      3&    -18.1&   0&    0&    0& S0(9)(cross)\\
   178&   -120.6&    408.5&    425.7&     - &     - 
&   - &    -14.1&   1&    0&    1& dE3,N\\
   179&   -120.7&   -192.1&    227.1&      908&       27
&      3&    -18.9&   0&    1&    0& Sa\\
   180&   -117.0&   -272.0&    296.3&     - &     - 
&   - &    -13.3&   0&    0&    1& dS0\\
   181&   -114.2&    178.8&    212.0&     1113&       53
&      2&    -14.1&   1&    0&    1& dE2,N\\
   182&   -112.2&     26.5&    115.3&     1669&       11
&      3&    -16.4&   0&    1&    0& Sa0\\
\end{tabular}
\end{minipage}\end{table*}
  
\begin{table*}\begin{minipage}{160mm}
\contcaption{}

\begin{tabular}{rrrrrrrrrrrl}
 FCC& X& Y& R& cz& err& ref& $M_B$& nuc& late& dwarf& morph\\
 & kpc& kpc& kpc& km/s& km/s& & mag& & & & \\
\\
   183&   -112.4&   -361.6&    378.9&     - &     - 
&   - &    -14.1&   1&    0&    1& dS0(7),N\\
   184&   -108.9&    -20.1&    110.8&     1257&       12
&      2&    -19.0&   0&    0&    0& SB0\\
   185&   -102.9&    200.7&    225.4&     - &     - 
&   - &    -11.8&   0&    0&    1& dE3\\
   186&   -103.7&   -937.0&    942.9&     - &     - 
&   - &    -13.5&   0&    0&    1& dE4\\
   187&   -100.7&    295.7&    312.3&     - &     - 
&   - &    -13.8&   0&    0&    1& dE5\\
   188&    -99.8&    -48.8&    111.1&     1046&       43
&      3&    -15.2&   1&    0&    1& dE0,N\\
   190&    -95.1&     89.1&    130.1&     1770&       10
&      3&    -17.8&   0&    0&    0& SB0\\
   191&    -93.6&     22.2&     96.2&     - &     - 
&   - &    -12.0&   0&    0&    1& dE3\\
   192&    -93.0&   -152.8&    179.0&     - &     - 
&   - &    -11.7&   0&    0&    1& dE0\\
   193&    -91.1&   -103.1&    137.7&      905&       10
&      3&    -18.5&   0&    0&    0& SB0(5)\\
   194&    -83.9&    -86.7&    120.7&     1237&       84
&      4&    -13.1&   0&    0&    1& dE3\\
   195&    -78.3&    192.1&    207.3&     1315&       69
&      4&    -14.6&   1&    0&    1& dE5,N\\
   196&    -64.9&   -132.2&    147.4&     1797&      129
&      4&    -13.6&   0&    0&    1& dE6\\
   197&    -57.0&     53.8&     78.3&     - &     - 
&   - &    -12.2&   0&    0&    1& dE3\\
   198&    -53.6&   -613.9&    616.2&     - &     - 
&   - &    -13.4&   0&    0&    1& dE5\\
   199&    -52.8&   -445.5&    448.7&     - &     - 
&   - &    -12.3&   0&    1&    1& ImV\\
   200&    -40.8&    198.4&    202.5&     1184&      110
&      4&    -13.9&   0&    0&    1& dE2\\
   201&    -40.9&   -638.6&    640.0&     1922&      126
&      2&    -14.6&   0&    0&    1& dE4\\
   202&    -26.6&      3.7&     26.9&      825&       20
&      3&    -16.0&   1&    0&    1& d:E6,N\\
   203&    -23.7&    325.3&    326.2&     1124&       16
&      3&    -15.8&   1&    0&    1& dE6,N\\
   204&    -18.6&    811.1&    811.3&     1364&       26
&      3&    -16.4&   1&    0&    1& dS0(8),N\\
   205&    -24.9&   -923.4&    923.7&     1450&       27
&      3&    -14.4&   1&    0&    1& dE1,N?\\
   206&    -17.9&   -642.1&    642.4&     1402&       20
&      2&    -15.5&   0&    0&    1& dE0pec\\
   207&    -11.6&    112.2&    112.9&     1403&       20
&      3&    -15.4&   1&    0&    1& dE2,N\\
   208&    -12.0&    -28.0&     30.5&     1720&       50
&      1&    -14.0&   1&    0&    1& dE2,N\\
   209&     -7.9&    625.0&    625.0&     - &     - 
&   - &    -12.6&   1&    0&    1& dE5,N\\
   210&    -11.5&   -214.8&    215.2&     - &     - 
&   - &    -12.6&   0&    0&    1& dE0\\
   211&     -8.9&     66.6&     67.1&     2260&       22
&      3&    -15.0&   1&    0&    1& d:E2,N\\
   212&     -9.4&   -336.2&    336.3&     - &     - 
&   - &    -13.7&   0&    0&    1& dE1?\\
   213&      0.3&     -0.6&      0.7&     1440&       19
&      3&    -20.7&   0&    0&    0& E0\\
   214&      9.0&   -133.9&    134.2&     - &     - 
&   - &    -12.3&   0&    0&    1& dE0\\
   215&     10.2&   -107.0&    107.5&     - &     - 
&   - &    -12.1&   1&    0&    1& dE,N?\\
   216&     12.0&   -386.7&    386.8&     - &     - 
&   - &    -12.1&   0&    0&    1& dE0\\
   217&     14.7&   -445.7&    445.9&     - &     - 
&   - &    -11.6&   0&    0&    1& dE/ImV\\
   218&     19.5&     64.4&     67.3&     - &     - 
&   - &    -12.8&   0&    0&    1& dE4\\
   219&     27.5&    -50.6&     57.5&     1919&       15
&      3&    -20.4&   0&    0&    0& E2\\
   220&     31.1&     74.6&     80.8&     - &     - 
&   - &    -11.8&   0&    0&    1& dE2\\
   221&     43.2&   -226.3&    230.5&     1724&       77
&      4&    -13.6&   1&    0&    1& dE4,N\\
   222&     52.7&     27.7&     59.5&      792&       26
&      3&    -15.7&   1&    0&    1& dE0,N\\
   223&     59.8&    -95.8&    113.0&      781&       62
&      3&    -15.1&   1&    0&    1& dE0,N\\
   224&     78.9&   1277.2&   1279.5&     - &     - 
&   - &    -14.4&   0&    0&    1& ImVordE\\
   225&     80.3&   -385.1&    393.5&     - &     - 
&   - &    -11.7&   0&    0&    1& dE4\\
   226&     96.8&    149.3&    177.8&     - &     - 
&   - &    -13.5&   0&    0&    1& dE?\\
   227&     96.2&    -25.2&     99.4&     - &     - 
&   - &    -12.0&   1&    0&    1& dE0,N?\\
   228&     97.8&     44.8&    107.5&     - &     - 
&   - &    -12.5&   0&    0&    1& dE0\\
   229&    102.0&    -74.0&    126.1&     - &     - 
&   - &    -12.2&   0&    0&    1& dE0\\
   230&    110.3&    241.7&    265.5&     1088&       30
&      3&    -14.1&   1&    0&    1& dE5,N\\
   231&    115.1&    447.8&    462.1&     - &     - 
&   - &    -12.9&   1&    0&    1& dE0,N?\\
   232&    121.0&   1093.0&   1099.5&     - &     - 
&   - &    -13.9&   1&    0&    1& dE7,N\\
   233&    114.1&   -273.4&    296.5&     - &     - 
&   - &    -11.7&   0&    1&    1& ImV\\
   234&    121.2&    350.5&    370.7&     - &     - 
&   - &    -14.1&   1&    0&    1& dE5,N\\
   235&    118.6&    -60.8&    133.4&     1974&       20
&      2&    -17.9&   0&    1&    1& ImIII\\
   236&    119.0&   -134.7&    180.0&     - &     - 
&   - &    -12.1&   0&    0&    1& dE2\\
   237&    133.6&    708.2&    720.4&     - &     - 
&   - &    -13.5&   0&    0&    1& dE3\\
   238&    126.4&   -378.6&    399.5&     - &     - 
&   - &    -12.6&   1&    0&    1& dE5,N\\
   239&    126.5&   -715.4&    726.8&     - &     - 
&   - &    -12.4&   0&    0&    1& dE5\\
   240&    146.7&   1314.3&   1322.1&     - &     - 
&   - &    -14.8&   0&    1&    1& ImIV\\
   241&    135.9&     60.9&    148.8&     - &     - 
&   - &    -14.7&   1&    0&    1& dE0,N\\
   242&    128.7&   -765.8&    776.9&     - &     - 
&   - &    -13.5&   0&    0&    1& dE5\\
   243&    138.0&   -366.1&    391.6&     1404&       45
&      2&    -14.8&   1&    0&    1& dE1,N\\
\end{tabular}
\end{minipage}\end{table*}
  
\begin{table*}\begin{minipage}{160mm}
\contcaption{}

\begin{tabular}{rrrrrrrrrrrl}
 FCC& X& Y& R& cz& err& ref& $M_B$& nuc& late& dwarf& morph\\
 & kpc& kpc& kpc& km/s& km/s& & mag& & & & \\
\\
   244&    143.5&   -149.2&    207.2&      831&      108
&      4&    -13.1&   0&    0&    1& dE6/ImIV?\\
   245&    148.7&    149.3&    210.4&     2187&       25
&      3&    -15.3&   1&    0&    1& dE0,N\\
   246&    151.2&   -234.1&    279.0&     - &     - 
&   - &    -12.2&   0&    0&    1& dE2\\
   247&    157.6&    -73.6&    174.1&     1097&      108
&      4&    -13.5&   0&    0&    1& dE3/Im?\\
   248&    158.5&   -143.3&    214.0&     - &     - 
&   - &    -12.6&   0&    0&    1& dE3\\
   249&    153.5&   -719.2&    735.9&     1533&       10
&      3&    -17.7&   0&    0&    0& E0\\
   250&    156.4&   -683.4&    701.5&     - &     - 
&   - &    -14.2&   0&    0&    1& dE1\\
   251&    167.5&    149.0&    223.9&     - &     - 
&   - &    -12.3&   0&    0&    1& dE0\\
   252&    166.9&   -104.0&    196.9&     1415&       35
&      2&    -15.3&   1&    0&    1& dE0,N\\
   253&    168.1&   -833.5&    850.8&     1677&       57
&      2&    -15.0&   1&    0&    1& dE5,N?\\
   254&    179.2&   -101.8&    206.4&     1517&       94
&      4&    -13.7&   1&    0&    1& dE0,N\\
   255&    186.9&    583.5&    612.2&     1271&       10
&      3&    -17.6&   1&    0&    0& S0(6),N\\
   256&    184.5&    173.1&    252.6&     - &     - 
&   - &    -11.2&   0&    0&    1& dE3\\
   258&    186.8&    -83.7&    205.0&     - &     - 
&   - &    -11.7&   0&    0&    1& dE2\\
   259&    187.6&    -22.5&    189.0&     - &     - 
&   - &    -13.5&   0&    0&    1& dE2\\
   260&    194.8&    101.9&    219.5&     1493&       59
&      3&    -14.3&   1&    0&    1& dE0,N\\
   261&    208.6&    586.8&    622.1&     1492&       42
&      2&    -15.5&   1&    0&    1& dE3pec,N/ImI\\
   262&    203.0&   -173.9&    267.8&     - &     - 
&   - &    -13.5&   0&    0&    1& dE1\\
   263&    218.8&    195.8&    293.0&     1733&        8
&      4&    -16.7&   0&    1&    0& SBcdIII\\
   264&    216.3&    -48.5&    221.8&     1888&       43
&      3&    -14.5&   1&    0&    1& dS0(8),N\\
   265&    234.9&    686.3&    724.5&     - &     - 
&   - &    -13.2&   0&    0&    1& dE5\\
   266&    228.6&     97.8&    248.3&     1551&       39
&      4&    -15.4&   1&    0&    1& dE0,N\\
   267&    237.6&    579.1&    625.1&      834&       10
&      2&    -15.3&   0&    1&    0& SmIV\\
   268&    234.5&   -411.8&    474.7&     - &     - 
&   - &    -13.0&   0&    0&    1& dE1\\
   269&    247.4&     55.2&    253.3&     - &     - 
&   - &    -13.0&   0&    0&    1& dE0\\
   270&    243.1&   -762.3&    801.2&     - &     - 
&   - &    -11.6&   0&    0&    1& dE\\
   271&    259.3&    209.7&    332.8&     - &     - 
&   - &    -12.6&   0&    0&    1& dE1\\
   272&    263.1&      2.3&    263.1&     - &     - 
&   - &    -12.1&   0&    0&    1& dE\\
   273&    272.0&    347.5&    440.2&     - &     - 
&   - &    -12.5&   0&    0&    1& dE2\\
   274&    270.2&    -31.5&    272.2&      950&       45
&      3&    -14.8&   1&    0&    1& dE0,N\\
   275&    272.1&    -38.6&    275.0&     - &     - 
&   - &    -12.0&   0&    0&    1& dE\\
   276&    273.1&     19.4&    273.7&     1382&       12
&      4&    -19.5&   0&    0&    0& E4\\
   277&    278.0&    103.4&    296.1&     1613&       25
&      4&    -17.5&   0&    0&    0& E5(boxy)\\
   278&    287.8&    551.4&    620.7&     2125&       30
&      2&    -14.5&   1&    0&    1& dE6,N\\
   279&    276.9&   -431.7&    514.1&     - &     - 
&   - &    -14.6&   1&    0&    1& dE0,N\\
   280&    291.9&   -176.5&    341.9&     - &     - 
&   - &    -12.7&   0&    0&    1& dE1\\
   281&    294.3&   -145.7&    329.1&     - &     - 
&   - &    -13.4&   0&    0&    1& dE1\\
   282&    309.6&    534.2&    616.0&     1251&       19
&      3&    -16.8&   0&    1&    1& ImIV/dEpec;emission\\
   283&    286.1&   -970.4&   1013.2&     - &     - 
&   - &    -13.5&   0&    0&    1& dE2\\
   284&    315.5&     37.4&    317.5&     - &     - 
&   - &    -12.3&   0&    0&    1& dE1\\
   285&    320.1&   -286.6&    430.9&      891&        6
&      2&    -17.1&   0&    1&    0& SdIII?\\
   286&    339.5&    281.8&    439.9&     1673&       82
&      3&    -13.2&   1&    0&    1& dE0,N?\\
   288&    354.5&    527.6&    633.9&     1088&       20
&      3&    -15.9&   1&    0&    1& dS0(9),N\\
   289&    351.8&    263.6&    438.3&     - &     - 
&   - &    -12.6&   0&    0&    1& dE0\\
   290&    363.2&   -141.2&    390.6&     1366&       12
&      3&    -18.5&   0&    1&    0& ScII\\
   291&    369.2&     81.9&    377.7&     - &     - 
&   - &    -11.9&   0&    0&    1& dE5\\
   292&    396.8&    840.9&    927.5&     - &     - 
&   - &    -14.2&   1&    0&    1& dE6,N?\\
   293&    419.9&   -141.8&    444.2&     - &     - 
&   - &    -13.7&   1&    0&    1& dE1,N\\
   294&    423.4&    -75.2&    430.5&     - &     - 
&   - &    -14.1&   1&    0&    1& dE1,N?\\
   295&    429.1&     94.9&    438.9&     - &     - 
&   - &    -12.4&   0&    0&    1& dE0\\
   296&    432.6&     88.8&    440.9&      856&       41
&      3&    -15.0&   1&    0&    1& dE1,N\\
   297&    435.8&   -185.8&    475.1&     - &     - 
&   - &    -13.5&   0&    0&    1& dE1\\
   298&    443.5&    -81.4&    451.5&     1620&       31
&      3&    -14.7&   1&    0&    1& dE2,N\\
   299&    453.3&   -504.2&    680.8&     2151&       40
&      2&    -14.2&   0&    1&    0& Sd(onedge)\\
   300&    458.2&   -303.4&    551.6&     - &     - 
&   - &    -15.2&   1&    0&    1& dE4,N\\
   301&    464.6&   -182.4&    500.5&     1038&       13
&      3&    -17.1&   0&    0&    0& E4\\
   302&    477.2&    -42.0&    479.3&      806&       23
&      2&    -15.6&   0&    1&    0& Sdm(onedge)\\
   303&    471.0&   -518.9&    703.8&     1980&       31
&      2&    -15.8&   1&    0&    1& dE1,N\\
   304&    505.6&    329.9&    601.3&     - &     - 
&   - &    -12.5&   0&    0&    1& dE1\\
\end{tabular}
\end{minipage}\end{table*}
  
\begin{table*}\begin{minipage}{160mm}
\contcaption{}

\begin{tabular}{rrrrrrrrrrrl}
 FCC& X& Y& R& cz& err& ref& $M_B$& nuc& late& dwarf& morph\\
 & kpc& kpc& kpc& km/s& km/s& & mag& & & & \\
\\
   305&    493.0&   -569.8&    756.9&     1228&       25
&      2&    -15.6&   0&    0&    1& dS0(6)\\
   306&    511.3&   -312.7&    601.8&      898&       14
&      2&    -15.7&   0&    1&    0& SBmIII\\
   307&    522.5&    136.1&    538.6&     - &     - 
&   - &    -13.6&   0&    0&    1& dE3\\
   308&    522.3&   -317.0&    613.5&     1497&        3
&      1&    -17.5&   0&    1&    0& Sd(onedge)\\
   309&    534.7&   -479.1&    721.4&     - &     - 
&   - &    -13.8&   0&    0&    1& dE2\\
   310&    542.1&   -435.1&    698.5&     1373&       13
&      2&    -17.8&   0&    0&    0& SB0\\
   312&    560.4&    176.9&    586.0&     1890&       22
&      3&    -17.8&   0&    1&    0& Scd(onedge)\\
   313&    579.4&    266.8&    635.4&     - &     - 
&   - &    -13.9&   0&    0&    1& dS0(9)\\
   314&    566.3&   -596.6&    826.9&     - &     - 
&   - &    -14.8&   1&    0&    1& dE2,N\\
   315&    624.0&    607.7&    866.4&     1071&       15
&      2&    -18.7&   0&    1&    0& Sab(rs)II\\
   316&    599.7&   -344.6&    695.0&     1546&      105
&      2&    -15.0&   1&    0&    1& dE3,N\\
   317&    636.1&    934.3&   1124.7&     - &     - 
&   - &    -13.6&   0&    0&    1& dE1\\
   318&    608.4&   -306.0&    684.0&     - &     - 
&   - &    -15.2&   1&    0&    1& dE2,N\\
   319&    648.3&   1099.0&   1270.0&     1445&       61
&      2&    -14.9&   1&    0&    1& dE6,N\\
   320&    671.0&   1115.5&   1295.6&     - &     - 
&   - &    -14.0&   1&    0&    1& dE5,N?\\
   321&    641.8&   -175.9&    667.4&     - &     - 
&   - &    -12.5&   0&    0&    1& dE\\
   322&    618.8&  -1093.1&   1262.6&      978&       15
&      2&    -18.8&   0&    1&    0& Sd\\
   323&    642.5&   -318.8&    720.6&     - &     - 
&   - &    -13.8&   0&    0&    1& dE2/ImV\\
   324&    659.4&   -356.5&    753.4&     1493&       44
&      2&    -16.0&   0&    0&    1& dS0(8)\\
   325&    682.8&    148.0&    696.9&     - &     - 
&   - &    -13.1&   0&    0&    1& dE3\\
   326&    670.4&   -452.4&    813.4&     - &     - 
&   - &    -12.7&   0&    0&    1& dE1\\
   327&    669.9&   -724.3&    992.8&     - &     - 
&   - &    -13.9&   0&    0&    1& dE3?\\
   328&    696.9&   -229.4&    736.4&     - &     - 
&   - &    -12.9&   0&    0&    1& dE\\
   329&    737.8&   -175.4&    760.6&     - &     - 
&   - &    -13.1&   0&    0&    1& dE\\
   330&    759.5&   -166.4&    779.7&     - &     - 
&   - &    -11.9&   0&    0&    1& dE\\
   331&    747.9&   -531.9&    923.8&     - &     - 
&   - &    -14.2&   0&    0&    1& dE\\
   332&    800.8&   -172.9&    821.6&     - &     - 
&   - &    -15.6&   0&    0&    0& EorS0\\
   334&    848.4&     70.6&    850.2&     - &     - 
&   - &    -12.7&   0&    0&    1& dE(boxy)\\
   335&    857.4&   -160.3&    874.7&     1367&       24
&      2&    -17.1&   0&    0&    0& E\\
   336&    884.0&     97.2&    887.8&     1956&       67
&      2&    -13.7&   0&    0&    1& dEpec\\
   337&    877.5&   -430.6&    983.6&     - &     - 
&   - &    -13.5&   0&    0&    1& dE/ImV\\
   338&    985.3&    691.8&   1194.5&     1562&       16
&      2&    -17.0&   0&    0&    0& S0\\
   339&    934.7&   -846.3&   1272.1&     - &     - 
&   - &    -13.8&   1&    0&    1& dE,N\\
   340&   1020.8&   -832.1&   1329.4&     - &     - 
&   - &    -13.1&   0&    0&    1& dE\\
   470&   -773.7&    -93.4&    780.6&      723&       79
&      2&    -13.8&   0&    1&    1& S/Im\\
   729&   -495.9&    135.6&    512.9&     1676&       31
&      2&    -14.8&   0&    0&    1& (d)SO\\
   904&   -326.7&    310.1&    449.2&     2254&       56
&      2&    -13.9&   0&    0&    1& (d)E\\
   905&   -325.3&    292.6&    436.3&     1243&       23
&      2&    -13.6&   0&    1&    1& ?emission\\
   934&   -957.1&    705.9&   1180.1&     1383&       85
&      4&    -12.7&   0&    0&    1& E\\
  1005&   -231.1&    994.2&   1019.8&     1256&       36
&      2&    -17.6&   0&    0&    0& SO\\
  1019&   -218.8&    738.7&    769.7&     2013&      135
&      4&    -13.0&   0&    0&    1& dS0\\
  1088&   -140.7&    759.0&    771.6&     1246&       33
&      4&    -12.4&   0&    1&    1& ?emission\\
  1108&   -120.5&    694.5&    704.6&     1734&       21
&      4&    -13.5&   0&    0&    1& (d)SO\\
  1241&    -14.5&    -20.0&     24.7&     2012&       91
&      4&    -16.6&   0&    0&    1& dE3\\
  1379&    104.8&    837.0&    843.4&      745&       21
&      2&    -14.0&   0&    0&    1& dE\\
  1554&    249.7&     35.7&    252.1&     1642&       52
&      2&    -13.6&   0&    0&    1& E\\
  2144&    841.7&   1114.0&   1387.0&     1184&       25
&      2&    -14.2&   0&    0&    1& (d)E\\
\end{tabular}\\

Notes: The catalogue number (FCC), projected position offsets and
radius from the central galaxy NGC~1399 (X, Y, R), absolute magnitude
($M_B$), and morphological type (``morph'') are all taken or
calculated from the Fornax Cluster Catalog \citep{Fer89}. The measured
radial velocities and uncertainties (cz; err) are from the following
references (ref):  1, the NED
database; 2, \citet{DGH01}; 3, Karick (PhD thesis 2003); 4,
\citet{DPJ00}. The binary flags for nucleated (nuc), late-type
(late) and dwarf galaxies have been derived from the FCC morphology,
as described in the text.
\end{minipage}\end{table*}


\begin{table}
\caption{The UCD sample.}
\begin{tabular}{rrrrrr}
 X& Y& R& cz& err& $M_B$\\
 kpc& kpc& kpc& km/s& km/s& mag\\
\\
     -8.6&      4.3&      9.7&     1398&       77&    -10.3\\
     -0.2&    -10.3&     10.3&     1460&       77&    -10.2\\
      9.3&     -6.6&     11.5&     1365&       56&    -10.5\\
      0.1&     11.7&     11.7&     1491&       73&    -10.6\\
      9.5&      7.4&     12.0&     1377&       85&    -10.3\\
     12.3&     -0.5&     12.3&     1445&       86&     -9.9\\
    -12.5&     -3.8&     13.0&     1332&       63&    -10.0\\
     13.3&     -0.2&     13.3&     1230&      112&    -10.5\\
     -8.5&     10.4&     13.4&     1357&      148&    -10.8\\
     -4.7&    -14.0&     14.7&     1158&       64&    -10.2\\
    -14.8&      4.0&     15.3&     1125&      101&    -10.8\\
    -17.5&      1.7&     17.6&     1377&       96&     -9.8\\
     16.8&     -5.9&     17.7&     1574&      117&    -10.3\\
     -2.7&    -18.1&     18.3&     1475&       72&    -10.2\\
    -21.6&      7.3&     22.8&     1702&       55&    -10.8\\
      0.1&     23.7&     23.7&     1607&       47&    -10.9\\
    -22.1&     16.9&     27.8&     1549&       64&    -11.2\\
    -26.9&    -11.4&     29.2&     1312&       57&    -12.5\\
    -28.3&     16.6&     32.9&     1212&       32&    -11.9\\
    -26.6&     23.0&     35.2&     1510&       64&    -10.8\\
    -13.5&    -35.1&     37.6&     1490&       69&    -10.5\\
     15.4&    -36.1&     39.3&     1980&       88&    -11.5\\
     -6.8&     40.7&     41.3&     1370&       64&    -10.6\\
     25.7&    -39.5&     47.1&     1744&       89&    -10.9\\
     19.9&    -43.3&     47.6&     1845&       87&    -10.7\\
     29.7&    -38.0&     48.3&     1591&       36&    -13.6\\
    -23.3&    -47.1&     52.6&     1764&       52&     -9.9\\
    -52.0&    -11.7&     53.3&     1641&       63&    -10.4\\
     57.7&      8.8&     58.4&     1022&       46&    -10.5\\
    -53.8&     24.1&     59.0&     1326&       82&    -11.2\\
    -54.5&     28.0&     61.3&     1146&       86&    -10.6\\
     -4.1&    -62.3&     62.4&     1698&       52&    -10.8\\
\end{tabular}

\end{table}

\begin{table}
\contcaption{}
\begin{tabular}{rrrrrr}
 X& Y& R& cz& err& $M_B$\\
 kpc& kpc& kpc& km/s& km/s& mag\\
\\
     21.9&    -59.4&     63.3&     1893&       68&    -10.2\\
    -60.2&     38.8&     71.6&     2226&       87&    -10.0\\
    -72.7&    -18.6&     75.1&     1828&       77&    -10.8\\
     61.2&     45.2&     76.0&     1420&       64&    -11.1\\
     -6.2&     76.7&     77.0&     1564&       73&    -11.1\\
    -20.1&    -75.3&     77.9&     1307&       63&    -10.1\\
     79.3&     -8.1&     79.7&     1920&       40&    -12.5\\
    -54.0&     69.0&     87.6&     1367&       71&    -12.4\\
     88.3&      0.2&     88.3&     1448&      101&    -11.3\\
    -75.8&    -53.2&     92.7&     1496&       55&    -11.0\\
    -55.7&    -83.5&    100.5&     1175&       61&    -10.0\\
     81.0&     67.8&    105.6&     1800&       93&    -10.4\\
   -115.3&    -21.1&    117.2&     1375&       46&    -10.0\\
   -101.3&    -64.3&    120.1&     1491&       39&    -11.5\\
   -120.0&    -15.1&    121.0&     1373&       92&    -10.4\\
   -136.0&     43.5&    142.7&     1817&      104&    -10.4\\
    131.8&     67.0&    147.7&     1618&       73&    -10.2\\
   -145.0&     29.0&    147.8&     1350&       97&    -11.2\\
   -149.9&    -55.6&    160.0&     1462&       76&    -10.9\\
   -144.1&     74.4&    162.0&     1297&       45&    -11.1\\
     99.5&    131.6&    164.8&     1355&       72&    -11.6\\
   -119.4&   -125.4&    173.4&     1340&       90&    -10.4\\
     77.5&   -155.5&    173.8&     1528&       73&    -10.6\\
    138.0&    118.6&    181.8&     1433&       59&    -10.8\\
    152.7&    164.9&    224.5&     1475&       57&    -10.0\\
   -143.4&    177.5&    227.9&     1658&       65&    -10.2\\
    220.2&   -162.6&    274.2&     1629&       56&    -10.3\\
   -256.8&    -99.1&    275.7&     1369&       68&    -10.3\\
\end{tabular}\\

Notes: The projected position offsets and radius from the central
galaxy NGC~1399 (X, Y, R), measured radial velocities and
uncertainties (cz; err), and absolute magnitudes ($M_B$), are all
taken or calculated from the Fornax Cluster Spectroscopic Survey
\citep{DPJ00}.

\end{table}

\section{Variation in line-of-sight velocity with radius.}
\label{sec:losvel}

This appendix calculates the expected variation in the line-of-sight
velocity dispersion of UCDs with radius resulting from anisotropic
motions in a declining density profile.

We take the velocity dispersion tensor to be aligned with the radial
direction and to have diagonal components $\sigma_{\rm r}$,
$\sigma_{\rm t}$ \& $\sigma_{\rm t}$, where $\sigma_{\rm r}$ and
$\sigma_{\rm t}$ are the radial and tangential components of velocity
dispersion, respectively.  The velocity anisotropy parameter is
defined as $\beta=1-\sigma_{\rm t}^2/\sigma_{\rm r}^2$---thus
$\beta=0$ for isotropic orbits and $\beta>0$ for preferentially radial
orbits.  We assume that both $\sigma_{\rm r}^2$ and $\beta$ are
constant throughout the cluster.  In practice one might expect some
radial variation in these quantities, but the data are insufficient to
constrain more complex models.

The line-of-sight velocity dispersion $\sigma_{\rm los}$ at a given
projected radius $R$ is then given by a density-weighted integral
along the line of sight:
\begin{eqnarray}
\sigma_{\rm los}^2(R) & = &
\frac{\int_0^\infty\rho(r)(\sigma_{\rm r}^2\sin^2\theta +
  \sigma_{\rm t}^2\cos^2\theta)dz}{\int_0^\infty\rho(r)dz} \\
& = & \sigma_{\rm r}^2 \left[ 1 - \beta
  \frac{\int_0^\infty\rho(r)\cos^2\theta\,dz}
       {\int_0^\infty\rho(r)\,dz} \right] \\
& = & \sigma_{\rm r}^2 \left[ 1 - \beta
  \frac{\int_0^{\pi/2}\rho(r)\,d\theta}
       {\int_0^{\pi/2}{\rho(r)\over\cos^2\theta}\,d\theta} \right], 
\label{eq:losvdisp}
\end{eqnarray}
where $z$ measures distance along the line of sight from the midpoint
through the cluster and $r$ is the radius to a point on that line such
that $z = R \tan\theta$ and $r = R / \cos\theta$.  The value of the
integral quotient in equation~\ref{eq:losvdisp} varies between
approximately 0.62 and 0.76 for inner and outer radial bins given in
Table~\ref{tab:sigmas} (the precise values depend upon the density
model for the UCDs but all acceptable fits to the data give similar
results).

\bsp

\label{lastpage}

\end{document}